\UseRawInputEncoding
\documentclass[a4paper,11pt]{article}
\pdfoutput=1
\usepackage{jheppub}
\usepackage{caption,subcaption}
\usepackage{graphicx} 
\usepackage{rotating,float,afterpage}
\usepackage{calrsfs}
\usepackage{slashed,cancel}
\usepackage{relsize}
\usepackage{amsmath,graphicx,epsfig,amssymb,multirow}
\RequirePackage{mathptmx,flushend}      
\usepackage[12hr]{datetime} 
\usepackage{orcidlink}
\usepackage{lineno}

\newcommand{\ttb}{t\bar{t}}
\newcommand{\bb}{b\bar{b}}

\newcommand{\ttHH}{t\bar{t}hh}
\newcommand{\tthh}{t\bar{t}hh}

\newcommand{\ttZH}{t\bar{t}Zh}
\newcommand{\ttZZ}{t\bar{t}ZZ}
\newcommand{\ttZ}{t\bar{t}Z}
\newcommand{\ttH}{t\bar{t}h}

\newcommand{\abinv}{ab$^{-1}$}

\title{Exploring Higgs EFT in $t\bar{t}hh$ at  High Luminosity LHC}
\author[a]{Ricardo D'Elia Matheus\textsuperscript{\orcidlink{https://orcid.org/0000-0003-2132-8251}},}
\affiliation[a]{Instituto de Física Teórica, Universidade Estadual Paulista (UNESP), 271 Dr. Bento Teobaldo Ferraz Street, São Paulo, Brazil}

\author[b]{Oscar J. P. Éboli\textsuperscript{\orcidlink{https://orcid.org/0000-0003-4107-6012}},}

\author[b]{Rafiqul Rahaman\textsuperscript{\orcidlink{https://orcid.org/0000-0002-3907-829X}}\footnote{Corresponding author},}
\affiliation{$^b$Departamento de Física Matemática, Instituto de Física, Universidade de São Paulo, Rua do Matão 1371, 05508-090, São Paulo, Brazil}

\author[c,d]{ and Aurore Savoy Navarro\textsuperscript{\orcidlink{https://orcid.org/0000-0002-9481-5168}}}
\affiliation[c]{IRFU - CEA, University Paris-Saclay, DPhP and CNRS/IN2P3, France }
\affiliation[d]{Now at: The Hubert Curien Pluridisciplinary Institute, IPHC, CNRS/IN2P3, France}

\emailAdd{ricardo.matheus@unesp.br}
\emailAdd{eboli@if.usp.br}
\emailAdd{rafiqul@if.usp.br}
\emailAdd{aurore.savoy.navarro@cern.ch}

\abstract{
	The non-resonant production of a Higgs boson pair in association with a top–antitop quark pair ($pp\to t\bar{t}hh$) has only recently begun to be explored at the Large Hadron Collider (LHC) and provides a unique and largely uncharted probe of the top–Higgs sector, offering complementary sensitivity to the Higgs self-coupling and higher-dimensional interactions beyond the Standard Model. In this work, we present a detailed study of this process within the framework of Higgs Effective Field Theory (HEFT) at the High-Luminosity LHC (HL-LHC). A comparative analysis is performed using a traditional cut-based approach in the single-lepton channel and a multivariate parametric boosted decision tree method in both single-lepton and dilepton final states. We derive one- and two-parameter limits at 95\% confidence level on the HEFT couplings $\delta\kappa_\lambda$,  $c_2$, $c_{2g}$, $c_{tg}$, and $c_{tg2}$. The projected bound on $\delta\kappa_\lambda$ is weaker than current experimental constraints from dedicated Higgs-pair measurement; however, this coupling plays a critical role in shaping the multidimensional allowed parameter space. For the remaining HEFT couplings, where no direct experimental limits currently exist, our results provide the first sensitivity projections in the $t\bar{t}hh$ channel. Overall, this study demonstrates the strong potential of the $t\bar{t} hh$ production process to probe extended Higgs and top-quark interactions beyond the Standard Model through the exploitation of the $t\bar{t}hh$ data at the HL-LHC.
}

\begin{document}

\maketitle

\section{Introduction}
\label{sec:introduction}
In the Standard Model (SM) of particle physics, the Higgs boson (h) couples to fermions with a Yukawa-type interaction whose strength is proportional to the fermion mass.
The Higgs coupling to the top quark, the heaviest known fermion, is particularly relevant both in the SM and in models of new physics beyond the SM (BSM). For example, due to its size, the top-Higgs coupling strongly affects the stability of the electroweak vacuum via virtual contributions and it may play a special role in the mechanism of electroweak symmetry breaking~\cite{Dobrescu:1997nm,Chivukula:1998wd}.
Therefore, probing the Higgs-top coupling is instrumental in testing the SM and constraining BSM models, which may predict different coupling strengths and additional structures. \smallskip

The associated production of a top quark-antiquark pair with two Higgs bosons ($\ttHH$) is a rare process in the SM that offers a unique opportunity to probe not only the Yukawa coupling of the top quark but also extensions of the SM that modify the couplings of top quarks, Higgs bosons and gluons.    Within the SM framework, the $\ttHH$ production is hard to observe due to its small production cross section and large backgrounds. Nevertheless, new physics effects can significantly increase the $\ttHH$ cross section, therefore, allowing the search for BSM physics. \smallskip

The  non-resonant $\ttHH$ production process has already begun to be studied in Runs 2 and 3 at the LHC.
Furthermore, substantial detector upgrades and an increase in integrated luminosity planned for the LHC High Luminosity Run (HL-LHC)~\cite{Contardo:2015bmq}, will provide an even much better environment for studying such a rare process, as stressed by a prospective study performed both within the SM and BSM frameworks ~\cite{CMS:2022rbr}.
Recent phenomenological studies have explored $\ttHH$ production to constrain the Higgs self-coupling and possible newphysics effects within effective field theory (EFT) and resonant composite Higgs scenarios at the HL-LHC and future 100~TeV proton–proton collider~\cite{Banerjee:2019jys,Li:2019wpa}.
Early theoretical explorations of this process were performed in Refs.~\cite{Englert:2014uqa,Liu:2014rva}, while the ATLAS experiment has assessed the SM sensitivity to $\ttHH$ production in the HL-LHC era~\cite{ATL-PHYS-PUB-2016-023}. CP-violating effects  using single- and double-Higgs production at the HL-LHC have also been studied in Ref.~\cite{Bhardwaj:2023ufl}.
This makes the present study quite timely. \smallskip

Many extensions of the SM predict not only the existence of new states, but also modifications of the Higgs boson interactions. For instance, in composite Higgs models,  the Higgs boson is a pseudo-Nambu Goldstone boson and the fermion sector of these models leads to changes to the Higgs couplings to top quarks. More specifically, the couplings $t\bar{t}h$ and $\tthh$ as well as the Higgs couplings to gluons differ from the SM ones~\cite{Bautista:2020mxw}.    \smallskip

EFTs provide a framework to perform model-independent searches for new physics whose characteristic mass scale is larger than the LHC energy. The eﬀective Lagrangian depends on the particle content as well as the symmetries at low energies. In this work, we do not assume that the observed Higgs boson belongs to a $SU(2)_L$ doublet, hence we perform our analyses in the Higgs effective field theory (HEFT) framework. HEFT provides a more general framework for describing electroweak symmetry breaking and the SM gauge symmetry is realized non-linearly. Furthermore, it is more general in the sense that it contains the scenario in which the Higgs boson is part of a $SU(2)_L$ doublet and possesses fewer relations among the new effective interactions.  \smallskip

The structure of this paper is as follows. Section~\ref{sec:theo} presents an overview of the theory context in which this  study is performed. The simulation of the signal and  the backgrounds are detailed in section~\ref{sec:simulation}. The potential for extracting the Standard Model $\tthh$ signal at HL-LHC is described in section~\ref{sec:sm-signal}. The way to probe the corresponding HEFT couplings in $\tthh$ process is addressed in section~\ref{sec:extract}. The main results and their interpretation are summarized in section~\ref{sec:results}.\smallskip 

\section{The theoretical context}
\label{sec:theo}

In the HEFT scenario, the physical Higgs boson ($h$) is treated as a singlet under the SM gauge symmetry while the three Goldstone bosons ($\vec{\pi}$) transform independently, unlike in the SM EFT (SMEFT) where they are components of an $SU(2)_L$ doublet. The Nambu-Goldstone bosons belong to  a dimensionless unitary matrix, in fact,  a bi-doublet~\cite{Appelquist:1980vg,Longhitano:1980tm,Feruglio:1992wf}: 
\begin{equation} 
	\mathbb{U}(x) = \exp\left( i \frac{ \sigma^a\pi^a(x)}{f} \right), \;\;\; \hbox{ with} \;\;\; \mathbb{U}(x) \to L \mathbb{U}(x) R^\dagger\;,
\end{equation} 
where $L$($R$) is the $SU(2)_{L(R)}$ transformation, allowing for a non-linear realization of electroweak symmetry breaking. Here, $\sigma^a$ are the Pauli  matrices  and $f$ is the scale associated to the Nambu-Goldstone bosons. Additional bosonic building blocks are the vector and scalar chiral fields
\begin{equation}
    \mathbb{V} = (D_\mu \mathbb{U}) \mathbb{U}^\dagger \;\;\; \hbox{ and } \;\;\; \mathbb{T} = \mathbb{U} \sigma^3 \mathbb{U}^\dagger \;,
\end{equation}
with $D_\mu$ standing for the covariant derivative. SM fermions are grouped into global $SU(2)_{L,R}$ doublets of the form
\begin{equation}
    Q_L \equiv \left( \begin{array}{c}
     u_L \\ d_L  
    \end{array}
    \right)
    \;\;\; \hbox{ and } \;\;\;
    Q_R \equiv \left( \begin{array}{c}
     u_R\\ d_R   
    \end{array}
    \right) \;.
\end{equation}
The HEFT Lagrangian is a power series constructed using the above building blocks and ordered using a chosen   power counting rule~\cite{Brivio:2016fzo, Buchalla:2022vjp, Brivio:2025yrr}.\smallskip

Here, we are interested in possible BSM effects in the $\ttHH$ process at the LHC; therefore, we focus on a subset of HEFT operators relevant to this channel. 
The  HEFT Lagrangian, restricted to operators directly relevant for $t\bar{t}hh$ production  is given by
~\cite{Carvalho:2015ttv,Cho:1994yu}:
\begin{eqnarray}\label{eq:eftLag}
	\mathcal{L}_{\rm{heft}} &=& \frac{1}{2} \partial_\mu h \, \partial^\mu h - \frac{1}{2} m_h^2 h^2 
	- (1 + \delta\kappa_\lambda) \lambda_{SM} v h^3
	 - \frac{m_t}{v}\left(  (1+\delta\kappa_t)h + \frac{c_2}{v} h h \right) 
	\left( \bar{t}_L t_R + \text{h.c.} \right) \nonumber \\ 
	&& +  \frac{\alpha_s}{12\pi v} 
	\left( c_g h - \frac{c_{2g}}{2v} hh \right) G_{\mu\nu}^a G^{a\mu\nu}  + \left\{ i \frac{g_s}{2v^2} \left( \bar{t} \sigma^{\mu\nu} T^a t_R \right)  G_{\mu\nu}^a h + \text{h.c.} \right\} c_{tg} \nonumber \\ 
&+&\left\{ i \frac{g_s}{2v^3} \left( \bar{t} \sigma^{\mu\nu} T^a t_R \right)  G_{\mu\nu}^a h h+ \text{h.c.} \right\} c_{tg2},	
\end{eqnarray}
where we omit those already tightly constrained such as modifications of the $gt\bar{t}$ interaction.
On one hand, the coefficients $\delta\kappa_\lambda$ and $\delta\kappa_t$ contribute to the tri-linear Higgs couplings and top-quark Yukawa couplings, respectively, that are already present in the SM at tree level. On the other hand, the couplings $c_2$, $c_{g}$, $c_{2g}$, $c_{tg}$ and $c_{tg2}$ do not exist in the SM, i.e., they are zero in the SM at tree level.   Notice that, in HEFT, the Wilson coefficients in Eq.~(\ref{eq:eftLag}) are independent in contrast to SMEFT where the linear realization of the SM gauge symmetry introduces relations between $\delta\kappa_t$ and $c_2$ as well as between $c_g$ and $c_{2g}$; see for instance Ref.~\cite{Azatov:2015oxa}. Nevertheless, it is possible to convert constraints on SMEFT into HEFT ones for a few channels. For instance, the SMEFT analysis of the gluon-gluon single Higgs  and $\ttH$ productions  can be recast into limits on $\delta\kappa_t$ and $c_{g}$, leading to strong bounds on these parameters. \smallskip

\begin{figure}[!ht]
\centering
\includegraphics[width=1\textwidth]{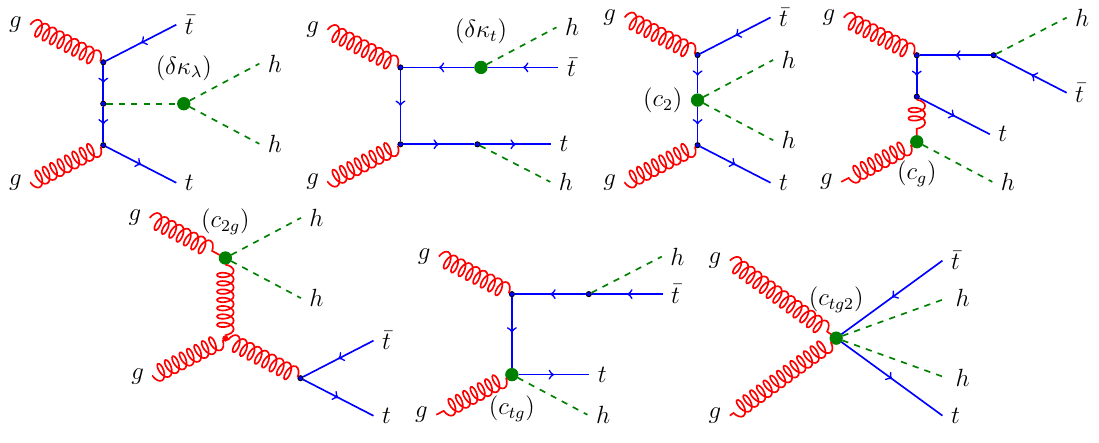}
\caption{\label{fig:FeynmanDiag} Representative Feynman diagrams for gluon-gluon initiated $t\bar{t}hh$ production with the HEFT couplings in Eq.~(\ref{eq:eftLag}) represented by shaded blobs.}
\end{figure}

In Figure~\ref{fig:FeynmanDiag}, the two first Feynman diagrams represent the SM couplings probed by $\ttHH$, namely the Higgs self-coupling and the Top-Higgs Yukawa coupling. The third diagram  involves the double Top-Higgs Yukawa coupling, labeled here as $c_2$, which is present at tree level in the $\ttHH$ channel,  contributes to double Higgs production at one loop, and is notably absent from $\ttH$ channel.  The other four following diagrams in this figure contain respectively the contribution of: $c_{g}$ for the $ggh$ coupling, $c_{2g}$ for the $gghh$ coupling, $c_{tg}$ for the $t\bar{t}g(g)h$ coupling, and $c_{tg2}$ for the $t\bar{t}g(g)hh$ coupling. The $c_{tg2}$ coefficient is particularly interesting as it reduces the
whole $gg \to t\bar{t}hh$ process down to a six point contact interaction. 
All these  couplings except $c_{tg2}$  also appear in interplay in other Higgs production processes at tree level, i.e., Higgs or double Higgs production as well as $\ttH$. The effective couplings $c_2$, $c_{g}$, $c_{2g}$, $c_{tg}$ and $c_{tg2}$ can be generated by some well motivated BSM models, such as the composite Higgs models.

\section{Signal and Background Simulation} 
\label{sec:simulation}

The SM non-resonant production process $\ttHH$ has a very small cross section of 0.981 fb  at next-to-leading order (NLO) in QCD~\cite{Frederix:2014hta} at the 14 TeV HL-LHC. Therefore, it is a rare process compared to a number of other Top-Higgs production processes. For example, the NLO production cross section for the $\ttH$ process is 612 fb~\cite{LHCHiggsCrossSectionWorkingGroup:2016ypw}, thus it is much larger than the one of the $\ttHH$ process. Therefore, we look for the most promising final states to study this channel. The  $\ttHH$ signal can be strengthened by requiring the two Higgs bosons to decay  into $b\bar{b}$ as well as  the two top quarks to decay into leptons, i.e., $\ttb\to\,2\ell$ and jets. We refer to this  topology as  the  di-lepton (DL) channel. \smallskip

Another clean topology is the single lepton (SL) channel where one of the top quarks decays leptonically ($t\to Wb, W\to\ell\nu_\ell$) and the other one decays hadronically ($t\to Wb, W\to q\bar{q'}$). This channel has a bit less clean signature than the DL one but with a larger production rate of more than a factor 3 compared to the DL channel. In our analyses,  the primary target is the SL channel, but we  also consider the additional contribution provided by the DL one.\smallskip

One of the main characteristics of the $\ttHH$ production is the large multiplicity of jets and b-jets, namely 6 $b$-jets in the DL and SL channels, with two additional light jets in the SL channel. Nevertheless,  the high multiplicity of jets tagged as b jets cannot be fully exploited to extract it from the huge SM backgrounds. Indeed, the requirement of six tagged $b$-jets, with a 60-70\% tagging efficiency~\cite{ATLAS:2014jfa} to be applied per jet dramatically reduces the overall signal efficiency. Hence, we concentrate on more inclusive topologies not requiring  the observation of all six $b$-jets but demanding that at least three tagged $b$-jets are present along with leptons and missing energy. We will also consider the possibility of demanding more than three tagged b-jets, depending on the analysis strategy, as explained and developed in Section~\ref{sec:sm-signal}.\smallskip

The dominant SM backgrounds are by far the $t\bar{t}$ + $b$-jets QCD production processes, such as $t\bar{t}+2b$  and $t\bar{t}+4b$, with large production cross section  of respectively $16.7$ pb and $373$ fb  at the HL-LHC. Furthermore, the $\ttb+2b2c$ and $\ttb+4c$ production processes can also provide leptons and $b$-jet with moderate ($>$10\%) $c$-quark to $b$-jet faking rate~\cite{CMS:DP2017013}. The $4t$ production process can further mimic the signal with a high multiplicity in light quark and $b$-quark jets, although it has a smaller cross section of  the order of $ 12$ fb. Additionally, the inclusive $\ttb$+jets process, despite containing only two $b$-quarks at the parton level, can, in fact,  produce many jets and $b$-jets due to its extremely large cross section, approximately $985$ pb at NNLO in QCD~\cite{ParticleDataGroup:2020ssz}.\smallskip

Among the electroweak (EW) dominated backgrounds, the production of a single Higgs along with $\ttb$, i.e. the $\ttH$ process, where $h\to\bb$, can provide a signature including leptons and tagged $b$-jets plus light jets and some missing energy, similar to the signal but with a production cross section of 612 fb. Likewise, the $\ttZ$ process, with 1018 fb NLO cross section~\cite{LHCHiggsCrossSectionWorkingGroup:2016ypw}, can provide leptons,  light quark jets and tagged $b$-jets in the final state, although $Z\to\bb$ decay has a smaller branching ratio ($\sim 15\%$~\cite{ParticleDataGroup:2018ovx}) than $h\to\bb$ decay ($60\%$~\cite{ParticleDataGroup:2018ovx}). \smallskip

There are further EW potential backgrounds. The  production processes $\ttZH$ and $\ttZZ$ provide event signatures very similar to the $\ttHH$ signal. Although these processes have, like $\ttHH$, very low cross sections of about 1.5 fb at LO for $\ttZH$ and 2.6 fb at NLO for $\ttZZ$, and thus provide a very low number of events, the experimentalists need sophisticated analysis tools to differentiate them from the signal. Likewise $\ttZZ$, but in a slightly minor way, $\ttb ZW$ presents some similarity to the SL and DL signal topology with leptons being misidentified. The processes $\ttH+2b$ and $\ttb V+2b$ with $V=Z/W$ with $~10$ fb cross section  also belong to the previous category with one on-shell Higgs or vector boson decaying into a $\bb$ pair.\smallskip

We simulate the signal and  background processes by generating events using \linebreak   \texttt{MadGraph5\_aMC@NLO} v3.5.6~\cite{Alwall:2014hca}  at LO in QCD. The heavy unstable particles are decayed using the \texttt{MadSpin}~\cite{Artoisenet:2012st} package 
to keep the possible spin-correlation effects among the final state decay products. The top quarks and $W$'s are decayed inclusively to all decay channels, while the Higgs and the $Z$ bosons are decayed to $\bb$ pair. The parton-level simulated events are then passed through \texttt{PYTHIA8}~\cite{Sjostrand:2014zea} for showering and hadronization with multi parton interaction (MPI). For the $\ttb$+jets samples which include up to two extra jets, we  implement the MLM matching scheme~\cite{Hoeche:2005vzu,Alwall:2007fs} with \texttt{PYTHIA8} to avoid double counting of jets with a 30 GeV of \texttt{xqcut} in \texttt{MadGraph5\_aMC@NLO}.\smallskip

Finally, a fast detector simulation is performed with \texttt{Delphes v3.5.0}~\cite{deFavereau:2013fsa} package with the default CMS card  with a few modifications related to the pseudo-rapidity ($\eta$), the radius parameter and the transverse momenta of the electrons, muons, and jets, to account for the considered HL-LHC scenario. The range in $|\eta|$ for the electrons, muons, and jets are extended to $3.0$  in the \texttt{Delphes} card, without changing the efficiency formulae in the last bin. The electrons and muons are reconstructed with a very small activity within the cone of  $\Delta R_{\ell}= \sqrt{\Delta \eta^2 + \Delta \phi^2}=0.2$ and minimum of $10$ GeV transverse momentum with other isolation parameter unchanged. The jets are reconstructed using \texttt{FastJet}~\cite{Cacciari:2011ma} package with the \texttt{anti-$k_t$} algorithm, with a radius parameter $R=0.4$ with a minimum transverse momentum $p_T=30$ GeV. The $b$-jets are identified using the default b-tagging algorithm present within the  \texttt{Delphes} card. \smallskip

\section{Features of the  $\ttHH$ Signal and Backgrounds}
\label{sec:sm-signal}

In this section, we study the main features of the $\ttHH$ signal as well as of its main backgrounds. We focus on the  jet and $b$-jet multiplicities  as well as other kinematic characteristics that can be explored to enhance the signal and deplete the backgrounds. \smallskip

In our basic selection of events, we required  at least 4 jets, being 3 b-tagged jets, within the rapidity interval $|\eta|\leq 3$ and transverse momentum $p_T$ in excess of 30 GeV. For the SL channel, we demanded exactly one lepton (electron or muon) with $|\eta| \leq 3$ and  $p_T\geq 25$ GeV and vetoed extra leptons with  $ p_T\geq 15$ GeV. Additionally, the SL events must possess missing transverse energy larger than  20 GeV.    In the DL channel, we required exactly two oppositely charged leptons  where the hardest lepton has $p_T\geq 25$ GeV and the second hardest lepton has $p_T\geq 15$ GeV. As the DL events exhibit more missing energy compared to SL, we imposed a MET cut of   40 GeV.  Our basic selection cuts are summarized in Table~\ref{tab:baselinecriteria}.\smallskip

\begin{table}[t!]
\centering
\footnotesize
\begin{tabular}{ll ll ll}
\hline
\multicolumn{2}{c}{Common Criteria} &
\multicolumn{2}{c}{SL} &
\multicolumn{2}{c}{DL} \\
\hline
Number of jets              & $\geq 4$        &
Number of leptons ($N_\ell$) & 1               &
Number of leptons ($N_\ell$) & 2 ($\ell^+\ell^-$) \\
Number of $b$-jets          & $\geq 3$        &
Lepton $p_T$                 & $\geq 25$ GeV   &
Leading lepton $p_T$         & $\geq 25$ GeV   \\
$p_T$ of all jets           & $\geq 30$ GeV       &
Second lepton $p_T$          & $<15$ GeV       &
Second lepton $p_T$          & $\geq 15$ GeV   \\
$|\eta|$ of all jets        & $\leq 3.0$          &
MET                          & $\geq 20$ GeV   &
MET                          & $\geq 40$ GeV   \\
$|\eta|$ of leptons         & $\leq 3.0$          &
                             &                 &
$\Delta R(\ell^+,\ell^-)$    & $\geq 0.2$      \\
\hline
\end{tabular}
\caption{\label{tab:baselinecriteria}Basic selection criteria for leptons, jets and MET after detector simulation.}	
\end{table}

We  estimate the expected number of events for the SM $\ttHH$ signal and the relevant backgrounds for integrated luminosity of 3~\abinv after applying the basic selection criteria in Table~\ref{tab:baselinecriteria} by taking into account detection efficiencies and production cross sections. The results are listed in Table~\ref{tab:sm-bsm-1l} where $V$ denotes an electroweak vector boson ($Z$, $W^\pm$) and $\ttb VV$ collectively refers to the $\ttb ZZ$, $\ttb ZW$ and $\ttb WW$ channels. The second column of this table provides the production cross sections taking into account the branching fractions $\mathrm{Br}(h\to\bb)=60\%$ and $\mathrm{Br}(Z\to\bb)=15.1\%$ wherever applicable, while the top quarks and $W$ bosons are treated inclusively in all decay modes. The third and fourth columns present the expected number of events at LO in QCD for the SL and DL channels, respectively, using the basic selection criteria for four jets of which three are $b$-tagged; we refer to this category as 4J3B category. In addition, we  also present the expected number of events for two more categories: 5J4B (five jets with four $b$-tags) and 6J5B (six jets with five $b$-tags), for the SL and DL channels.\smallskip

Examining Table~\ref{tab:sm-bsm-1l}, we  see that the largest background is $\ttb+\text{jets}$ followed by $\ttb+2b$ in all categories, as expected due to their large production cross sections. However, for higher jet and $b$-jet multiplicities, the background rejection becomes stronger for these two backgrounds compared to the others. For example, when going from 4J3B to 6J5B in the SL channel, the rejection of $\ttb+\text{jets}$ is about $99.5\%$, while that of $\ttH$ is $96.4\%$. The signal rejection, on the other hand, is about $79\%$, which is significantly lower than the corresponding rejections for the backgrounds $\ttb+\rm{jets}$ and $\ttH$ mentioned above. \smallskip

In Table~\ref{tab:sm-bsm-1l}, the signal significances for the SM $\ttHH$ process are presented at LO  for an integrated luminosity of 3 ab$^{-1}$. We also evaluate the significances at NLO by applying the NLO-to-LO $k$-factors, and NNLO/LO for $\ttb+{\rm jets}$ (see Table~\ref{tab:kfactor}), to the LO event yields; these results are also included in the table. In Table~\ref{tab:kfactor}, for processes with $4b$ and $2b2c$ final states, we assume a $k$-factor of 1.2  as higher-order corrections for these processes are not currently available in the literature.
We find that the significance for the SM $\ttHH$ process is very small, primarily due to its tiny production cross section. For example, the  significance in the SL  channels for the 6J5B category is  about 0.1 at LO, and it decreases further to 0.08 after incorporating the NLO $k$-factors.

\begin{table}[htb!]
	\centering
	\renewcommand{\arraystretch}{1.2}
	\setlength{\tabcolsep}{5pt}
 \small
 \begin{tabular}{llllllll}
\hline
	&  & \multicolumn{6}{c}{Number of Events}\\ \cline{3-8}
Process & $\sigma$ [fb] & \multicolumn{2}{c}{4J3B (basic-selection)}  &\multicolumn{2}{c}{5J4B} & \multicolumn{2}{c}{6J5B} \\\hline
&& SL & DL& SL & DL& SL & DL\\ \cline{3-4} \cline{5-6}  \cline{7-8}
$t\bar{t}hh$ ($S$)      & 0.33     & $1.54\times10^{2}$   & $1.5\times10^{1}$    & $8.7\times10^{1}$    & 7     & $3.2\times10^{1}$    & 2 \\ \hline
$t\bar{t}h$         & 287.9   & $7.33\times10^{4}$ & $5\times10^{3}$ & $1.99\times10^{4}$ & $9.34\times10^{2}$ & $2.63\times10^{3}$ & $8.1\times10^{1}$ \\
$t\bar{t}hZ$        & 0.14    & $6.4\times10^{1}$ & $6$ & $3.5\times10^{1}$ & $3$ & $1.3\times10^{1}$ & $1$ \\
$t\bar{t}Z$         & 106.6   & $2.3\times10^{4}$ & $1.85\times10^{3}$ & $5.93\times10^{3}$ & $3.44\times10^{2}$ & $7.35\times10^{2}$ & $2.7\times10^{1}$ \\
$t\bar{t}VV$        & 10.38   & $1.15\times10^{3}$ & $1.67\times10^{2}$ & $1.98\times10^{2}$ & $2\times10^{1}$ & $2.5\times10^{1}$ & $2$ \\
$t\bar{t}t\bar{t}$  & 11.83   & $5.7\times10^{3}$ & $1.08\times10^{3}$ & $2.32\times10^{3}$ & $3.89\times10^{2}$ & $5.86\times10^{2}$ & $7.9\times10^{1}$ \\
$t\bar{t}+4b$       & 373.3   & $6.83\times10^{4}$ & $5.44\times10^{3}$ & $2.1\times10^{4}$ & $1.45\times10^{3}$ & $4.51\times10^{3}$ & $2.58\times10^{2}$ \\
$t\bar{t}+2b2c$     & 105.5   & $1.9\times10^{4}$ & $1.66\times10^{3}$ & $5.49\times10^{3}$ & $4.09\times10^{2}$ & $1.07\times10^{3}$ & $6.6\times10^{1}$ \\
$t\bar{t}V+2b$      & 8.22    & $1.94\times10^{3}$ & $1.86\times10^{2}$ & $6.76\times10^{2}$ & $5.3\times10^{1}$ & $1.64\times10^{2}$ & $1.1\times10^{1}$ \\
$t\bar{t}h+2b$      & 9.38    & $3.35\times10^{3}$ & $3.1\times10^{2}$ & $1.48\times10^{3}$ & $1.14\times10^{2}$ & $4.35\times10^{2}$ & $2.7\times10^{1}$ \\
$t\bar{t}+4c$       & 7.81    & $1.07\times10^{3}$ & $9.3\times10^{1}$ & $2.54\times10^{2}$ & $1.9\times10^{1}$ & $4\times10^{1}$ & $2$ \\
$t\bar{t}+2b$       & $1.67\times10^{4}$ & $1.68\times10^{6}$ & $1.18\times10^{5}$ & $3.22\times10^{5}$ & $1.77\times10^{4}$ & $3.93\times10^{4}$ & $1.8\times10^{3}$ \\
$t\bar{t}+\mathrm{jets}$ & $6.11\times10^{5}$ & $1.2\times10^{7}$ & $4.66\times10^{5}$ & $8.68\times10^{5}$ & $2.87\times10^{4}$ & $6.02\times10^{4}$ & $1.7\times10^{3}$ \\
\hline
Tot. BKG ($B$)      & $6.29\times10^{5}$ & $1.4\times10^{7}$ & $5.99\times10^{5}$ & $1.25\times10^{6}$ & $5.01\times10^{4}$ & $1.1\times10^{5}$ & $4.06\times10^{3}$ \\
\hline
\multicolumn{2}{l}{LO Significance ($S/\sqrt{B}$)} & 0.04 & 0.02 & 0.08 & 0.03 & 0.10 & 0.03 \\ \hline
\multicolumn{2}{l}{NLO Significance} & 0.04 & 0.02 & 0.07 & 0.03&  0.08 & 	0.03 \\ \cline{3-8}
\hline
\end{tabular}
 \caption{\label{tab:sm-bsm-1l}	Expected number of events for the $\ttHH$ signal and background processes at LO in QCD for different jet and $b$-jet multiplicities in the SL and DL channels after applying the basic selection criteria (see Table~\ref{tab:baselinecriteria}) at the 14 TeV LHC with an integrated luminosity of 3 ab$^{-1}$. The cross sections ($\sigma$) given in the second column include the branching fractions $\mathrm{Br}(h\to\bb)=60\%$ and $\mathrm{Br}(Z\to\bb)=15.1\%$ wherever applicable, while the top quarks and the $W$ bosons are treated inclusively in all their decay modes. 
	The signal significances ($S/\sqrt{B}$) are shown in the bottom rows at LO, as well as after incorporating NLO-to-LO $k$-factors for the processes listed in Table~\ref{tab:kfactor}.}
\end{table}
\begin{table}[h!]
    \centering
    \renewcommand{\arraystretch}{1.2}
	\begin{tabular}{lc}
		\hline
		Process &  $k$-factor \\
		\hline
		$t\bar{t}hh$        & 1.07 [NLO/LO]~\cite{Frederix:2014hta}\\
		$t\bar{t}h$         & 1.28 [NLO/LO]~\cite{LHCHiggsCrossSectionWorkingGroup:2016ypw}\\
		$t\bar{t}Z$         & 1.44 [NLO/LO]~\cite{LHCHiggsCrossSectionWorkingGroup:2016ypw}\\
		$t\bar{t}+4b$       & 1.20 \\
		$t\bar{t}+2b2c$     & 1.20 \\
		$t\bar{t}+2b$       & 1.8 [NLO/LO]~\cite{Frederix:2014hta} \\
		$t\bar{t}$+jets     & 1.61 [NNLO/LO]~\cite{ParticleDataGroup:2020ssz}\\
		\hline
	\end{tabular}
\caption{\label{tab:kfactor} Processes with $k$-factors (N(N)LO/LO in QCD) larger than unity that are used in our analysis.}	
\end{table}

\begin{figure}[tb!]
	\centering	
	\includegraphics[width=0.325\textwidth]{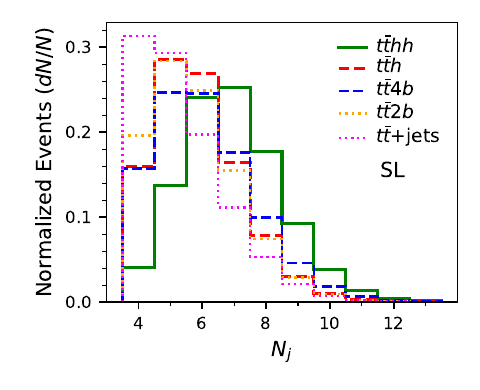}
	\includegraphics[width=0.325\textwidth]{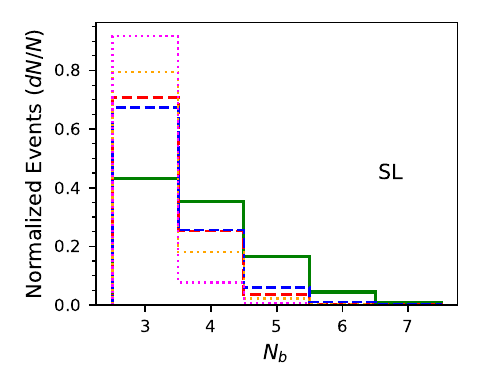}
	\includegraphics[width=0.325\textwidth]{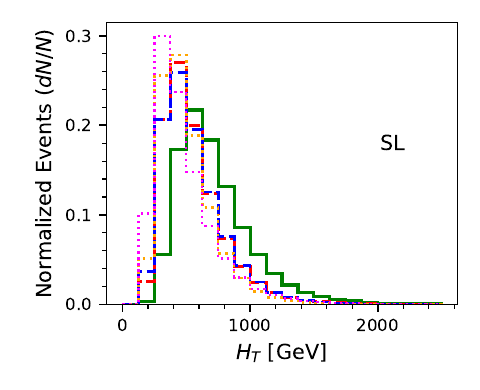}
	\includegraphics[width=0.325\textwidth]{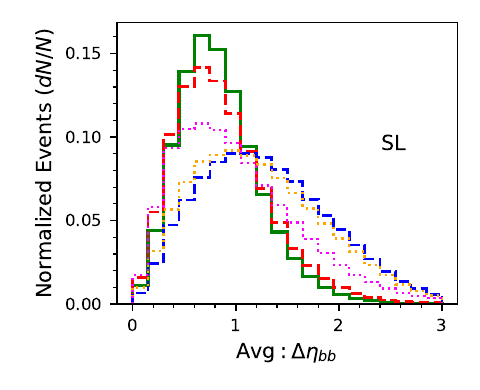}
	\includegraphics[width=0.325\textwidth]{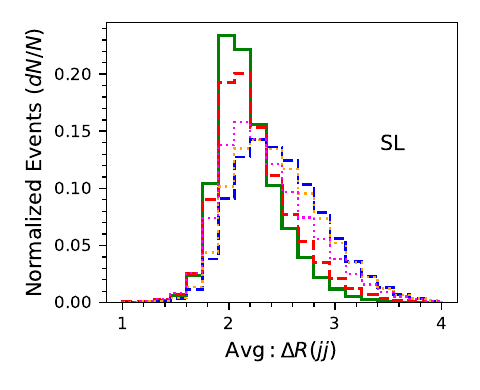}
	\includegraphics[width=0.325\textwidth]{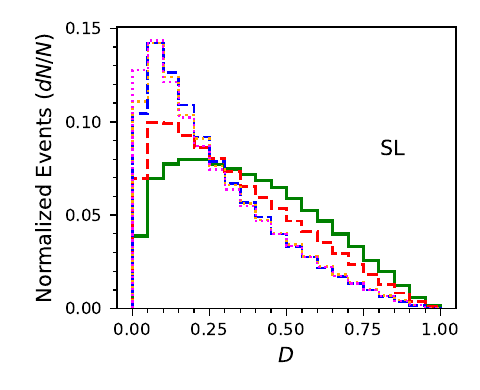}
	\caption{\label{fig:dist-sm-bkg} Normalized distributions for some sensitive variables are shown, for the four main backgrounds  $t\bar{t}h$, $t\bar{t}+4b$, $t\bar{t}+2b$ and $t\bar{t}+jets$) and the SM signal $t\bar{t}hh$ processes. 
    }
\end{figure}

After our basic selection, the background is much larger than the $\ttHH$ signal; therefore we analyzed various normalized kinematic distributions in order to boost the HL-LHC potential to probe the $\ttHH$ production. We depict in Fig.~\ref{fig:dist-sm-bkg} several kinematical distributions -- that is, number of jets and b-tagged ones, $H_T$, average rapidity separation between jets (b-tagged jets) $\Delta R(jj)$ ($\Delta\eta_{bb}$), and the event-shape variable $D$~\cite{Dasgupta:2003iq}  for the $\ttHH$ signal and the leading backgrounds. As expected, the $\ttHH$ signal contains higher jet and $b$-jet multiplicities ($N_j$ and $N_b$) than the largest backgrounds. In addition, the total transverse energy ($H_T$) is larger in signal events than in background ones. Furthermore, the average distance in rapidity of the (b-tagged) jets is concentrated at smaller values for the signal while the background distribution is wider. The kinematic variable $D$ also presents a different behavior between the signal and background events. \smallskip

Using the distributions of these variables, together with many others, we perform a BDT-based multi-class classification of the signal and background processes in the hope of improving the signal significance for the $\ttHH$ process\footnote{ The definition of the variables used in BDT is detailed in \autoref{sec:inputs}.}. After performing the multi-class BDT analysis using events that meet the basic selection criteria (i.e., those in the 4J3B category), the significance of the SM $\ttHH$ process increases  by a factor of approximately 1.5 compared to the corresponding values reported in Table~\ref{tab:sm-bsm-1l}.

Although observing the SM $\ttHH$ signal at the HL-LHC will be extremely hard, the process may become accessible if its rate is enhanced by possible BSM effects. For example, with certain choices of the HEFT couplings, the $\ttHH$ cross section can increase significantly. Therefore, by studying the $\ttHH$ process with HEFT couplings and comparing it with the relevant backgrounds, we can estimate the allowed ranges of these couplings; this is the main goal of this paper. In the next section, we extract the allowed ranges or limits on the HEFT couplings (see Eq.~(\ref{eq:eftLag})) using both a cut-based approach and a multi-class classification strategy based on Parametric BDTs. \smallskip

\section{Constraining the HEFT couplings}
\label{sec:extract}

In this section, we probe the HEFT couplings in the $\ttHH$ process and extract constraints using both cut-based and multivariate approaches, employing various kinematic observables. \smallskip

\subsection{Analysis strategy}
\label{subsec:strategy}

The analysis is based on MC simulations of $\ttHH$ events generated for a set of benchmark points (BPs) corresponding to different choices of HEFT couplings. These benchmark points, defined later in the text, span representative regions of the parameter space and allow us to study the impact of anomalous couplings on both kinematic distributions and overall event yields. To obtain constraints on the HEFT couplings, we follow two complementary strategies:

\paragraph{A. Cut-based approach.}
The cut-based strategy exploits characteristic features of various kinematic distributions in $\ttHH$ production. Starting from the baseline selection in the 4J3B category, additional cuts are applied to relevant kinematic variables to improve signal-to-background discrimination. In particular, the binned distribution of the scalar sum of jet transverse momenta, $H_T$, serves as a key observable. These optimized selections are designed to suppress dominant backgrounds such as $\ttb + \mathrm{jets}$ while retaining a significant fraction of the HEFT-sensitive $\ttHH$ signal.

\paragraph{B. Parametric BDT-based approach.}
In this strategy, we employ a multivariate method using a Parametric BDT (PM:BDT) architecture to maximize sensitivity to deviations induced by HEFT couplings. Rather than relying on hard threshold cuts, the PM:BDT exploits multidimensional correlations among several kinematic variables characteristic of $\ttHH$ production, including those sensitive to top and Higgs reconstruction. Events satisfying the baseline selection are used to train the PM:BDT simultaneously on multiple HEFT benchmark points, enabling interpolation across parameter space and providing smooth and continuous constraints on the couplings.

\subsection{HEFT choices for $\ttHH$ samples}
\label{subsec:HEFT-samples}

A global SMEFT analysis by CMS~\cite{CMS:2025ugn} constrains the dimension-six operator $c_{tH}$, which modifies the top Yukawa coupling, to the range $[-3.8,\,0.54]~\mathrm{TeV}^{-2}$, corresponding to $\delta\kappa_t \in [-0.23,\,0.03]$ in our HEFT parametrization. 
The same CMS global analysis~\cite{CMS:2025ugn} also bounds the Higgs--gluon operator to
$c_{HG}/\Lambda^2 \in [-0.004,\,0.008]~\mathrm{TeV}^{-2}$, which translates into HEFT bounds
$c_g \in [-0.08,\,0.15]$. These constraints force $\delta\kappa_t$ and $c_g$ so close to their SM values that their variation in the allowed region would have little effect in modifying the HEFT $\ttHH$ signal. We therefore choose to keep these two couplings fixed at zero, focusing our analysis on $\delta\kappa_\lambda$, $c_2$, $c_{2g}$, $c_{tg}$ and $c_{tg2}$.  

\smallskip

In order to extract the limits of these HEFT couplings, we simulate the $\ttHH$ process by choosing values of the couplings, varying one parameter at a time and two parameters at a time while keeping the others to their SM values, based on ranges of the SM signal and background cross sections. 
The ranges for the single parameter variations are determined such that the $t\bar{t}hh$ signal achieves a $3\sigma$ significance when the HEFT couplings are included, that is,  the signal must be $\sim 30$ times larger than the SM signal in the 6J5B category, see Table~\ref{tab:sm-bsm-1l}.\smallskip

We note that although these HEFT benchmark choices may appear to raise concerns regarding perturbativity or unitarity due to the energy-growing behavior of the partonic amplitudes, in practice the impact of this region at the LHC is strongly reduced by the rapidly falling gluon parton luminosities. Consequently, the signal is dominated by moderate-energy events, as also reflected in the steeply falling high-energy tails of the jet-invariant-mass and $H_T$ distributions shown in Fig.~\ref{fig:dist-sm-bkg}.

\begin{figure}[tb!]
\centering
\includegraphics[width=0.43\textwidth]{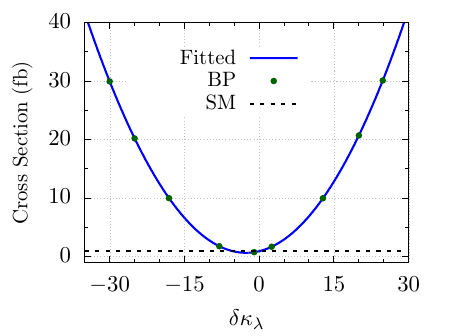}
\includegraphics[width=0.43\textwidth]{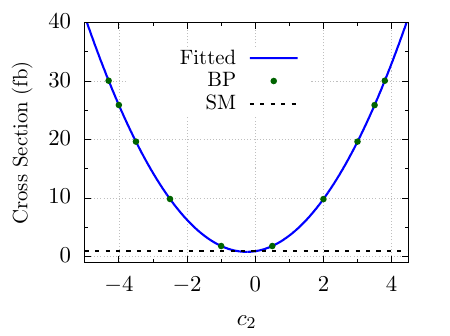}
\includegraphics[width=0.43\textwidth]{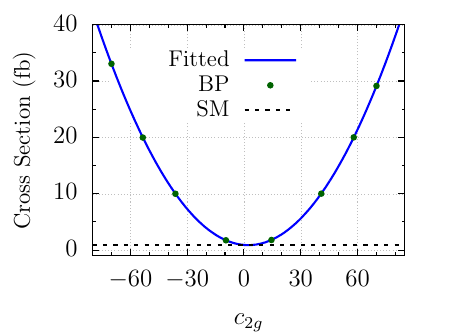}
\includegraphics[width=0.43\textwidth]{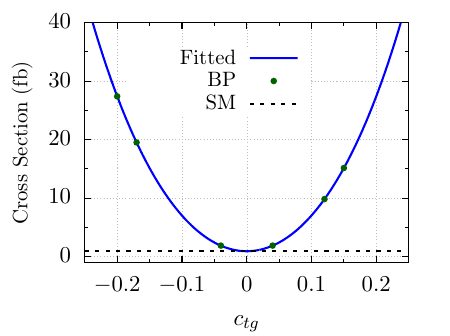}
\includegraphics[width=0.43\textwidth]{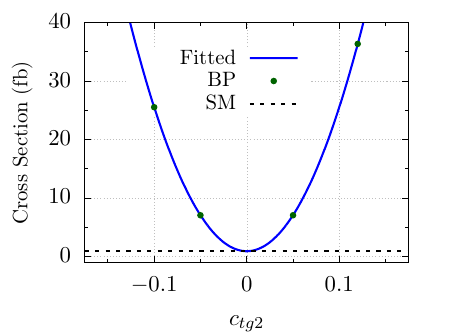}
\caption{\label{fig:prod-cs-one} Production cross section of $t\bar{t}hh$  at LO in QCD as a function of HEFT couplings  as given  in Eq.~(\ref{eq:prod-cs-HEFT}). The lines represent the fitted functions, whereas the blobs on the curve represent the benchmark points (BPs) where simulations are performed. The SM prediction is represented by the horizontal dashed black lines.}
\end{figure} 

The leading order production cross section of the $t\bar{t}hh$ process at the 14 TeV LHC is shown in Fig.~\ref{fig:prod-cs-one} as a function of the HEFT couplings with one parameter varied at a time. The continuous blue lines, green dots and dashed black lines represent respectively the fitted functions, the simulation of the  benchmark points and the SM value. 
The fitted function for the production cross sections (at LO) is given by
\begin{eqnarray} \label{eq:prod-cs-HEFT}
 \sigma_{\ttHH}(c_{\rm HEFT}) &= & \underbrace{0.914}_{\text{SM}} - 0.207 \times \delta\kappa_\lambda + 0.0391 \times \delta\kappa_\lambda ^2 \notag\\
    &-& 0.891\times c_2 + 1.79 \times c_2^2 \notag\\
 %
 %
    &+& 0.0281 \times c_{2g} + 0.0062 \times c_{2g}^2 ~\notag\\
    &-& 0.0052 \times c_{tg} + 590 \times c_{tg}^2 +0.0833 \times c_{tg}^3 + 1781 \times c_{tg}^4 \notag\\
    &+& 0.0212 \times c_{tg2} + 2462\times c_{tg2}^2  ~~~\rm{fb}, 
\end{eqnarray}
when one of the couplings is assumed non-zero at a time. As some diagrams can contain more than one power of the BSM couplings, each coupling is expanded up to the power that it appears at tree level in the cross section of the $\ttHH$ channel. In order to understand the interplay between the Wilson coefficients we also performed two-parameter analyses, that is, we also considered two non-vanishing parameters at the same time. For the two-parameter case, we choose some  10 to 16 benchmark points on the HEFT couplings.  The fitted function describing the simultaneous variation of two parameters is given in Eq.~(\ref{eq:sig-2param}) in Appendix~\ref{sec:sig-2param}  for completeness. \smallskip

In sections \ref{subsec:cut-based} and \ref{subsec:pm-bdt}, when studying the HEFT contributions to the aforementioned SL and DL channels, we have to  include the contribution coming from the modification of the $\ttH$ production, as it also decays into single or double leptons accompanied by jets. At LO the production of $\ttH$ is affected only by one of the  five couplings under study, $c_{tg}$,  and the fitted function analogous to Eq.~(\ref{eq:prod-cs-HEFT}) is:
\begin{equation}
\sigma_{\ttH}(c_{tg})=\underbrace{480}_{\text{SM}}-0.962\times c_{tg}+3.96\times10^4\times c_{tg}^2 ~~~\rm{fb}.
\end{equation}
Note that $c_{tg2}$ does not contribute to the $\ttH$ process.

\subsection{Constraints from cut-based analysis}
\label{subsec:cut-based}

\begin{figure}[htb!]
    \centering 
    \includegraphics[width=0.325\textwidth]{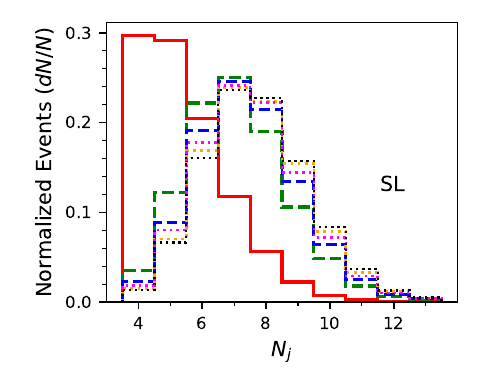}
    \includegraphics[width=0.325\textwidth]{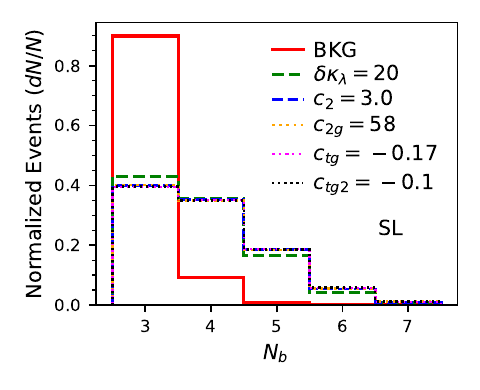}
    \includegraphics[width=0.325\textwidth]{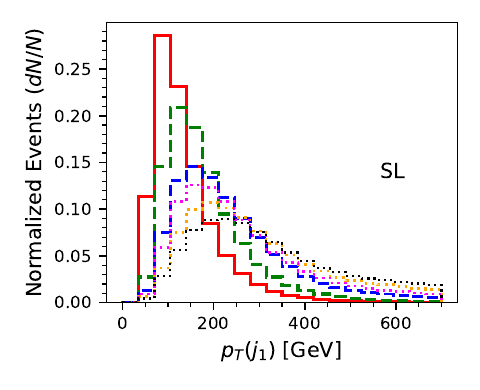}
    \includegraphics[width=0.325\textwidth]{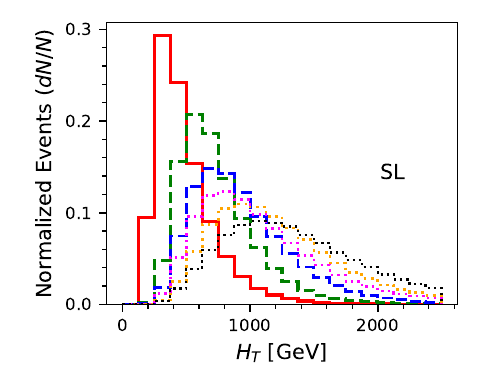}
    \includegraphics[width=0.325\textwidth]{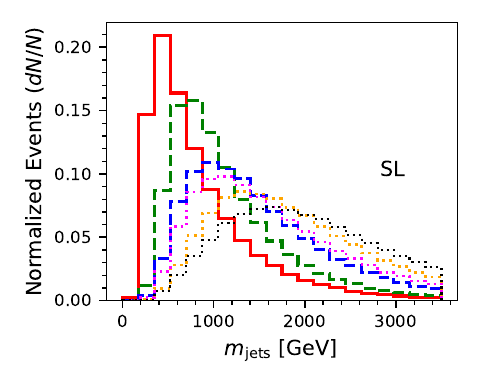}
    \includegraphics[width=0.325\textwidth]{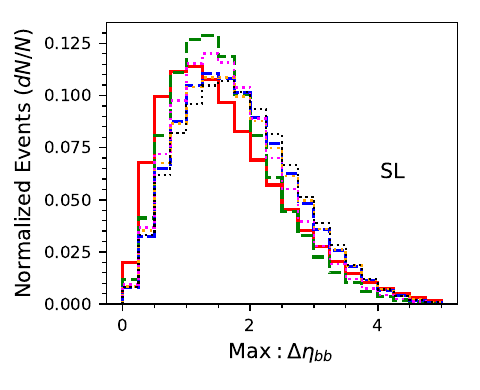}
    \includegraphics[width=0.325\textwidth]{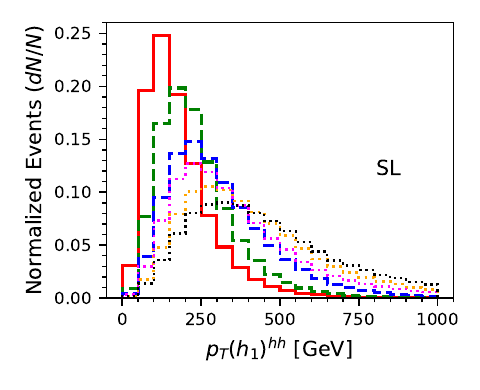}
    \includegraphics[width=0.325\textwidth]{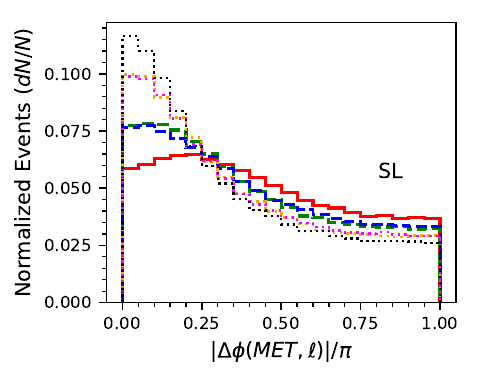}
    \includegraphics[width=0.325\textwidth]{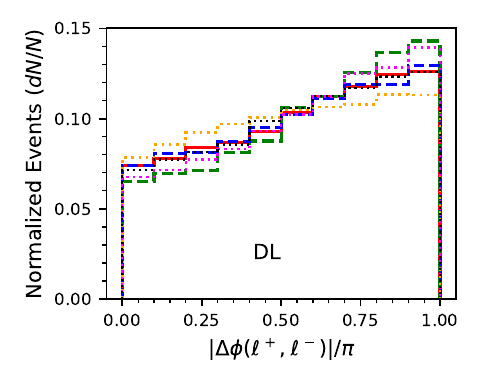}
    \caption{\label{fig:dist-bsm}  Normalized distributions for some HEFT-sensitive variables are shown for the combined background (BKG) and a few  selected benchmark points.}
\end{figure}

We study various kinematical distributions characteristic of the $\ttHH$ topology in addition to jet and $b$-jet multiplicities obtained from the baseline selection in the 4J3B category. 
We have already seen that the jet and $b$-jet multiplicities differ significantly between the SM $\ttHH$ signal and the backgrounds, and the inclusion of HEFT contributions further enhances these differences. In Fig.~\ref{fig:dist-bsm}, several HEFT-sensitive normalized distributions are shown for the combined background (BKG) and some selected reference points
\begin{equation}\label{eq:bp-dist}
\delta\kappa_\lambda=20,~c_2=3,~c_{2g}=58,~c_{tg}=-0.17,~c_{tg2}=-0.1 \;,
\end{equation} 
where we are assuming just one non-vanishing Wilson coefficient. These benchmark HEFT couplings are chosen such that they yield roughly $2\sigma$ sensitivity based on the  SM signal significance shown in  Table~\ref{tab:sm-bsm-1l}, corresponding to about 20 fb of production cross section for the $\ttHH$ process; see Fig.~\ref{fig:prod-cs-one}. Notice that the anomalous coupling $c_{tg}$ contributes at tree level to the $t\bar{t}h$ production, ergo we also add this contribution to the signal.\smallskip

As expected, the HEFT benchmark signals show a higher content of jets ($N_j$) and $b$-jets. Since the HEFT couplings modify the energy  dependence of the cross section, the jet $p_T$ and  $H_T$ as well as  the invariant mass of all jets ($m_{\rm jets}$) peak at larger values compared to the background. Additionally, reconstructed observables relevant to the $\ttHH$ topology, discussed in the  next paragraph, for example, the $p_T$ of the hardest reconstructed Higgs, show considerable separation from the background. Furthermore, angular variables such as the $\Delta\phi$ between the lepton and MET, as well as that between the lepton and the hardest reconstructed Higgs, also exhibit noticeable differences between the signal and background. \smallskip

In  addition to the above kinematic variables, we also reconstruct the di-boson final states $hh$, $ZZ$, and $hZ$, where each boson is assumed to decay into a pair of $b$-quarks. The reconstruction is performed by assigning the most likely combinations of jets to the bosons through a $\chi^{2}$-minimization procedure. For each di-boson hypothesis $[X,Y] = [h,h], [h,Z], [Z,Z]$, we define the corresponding variable 
$\chi^{2}$ as in Ref.~\cite{CMS:2022rbr} 
\begin{equation}
\chi^{2}_{XY}
=
\left( \frac{m_{j_{1} j_{2}} - m_{X}}{\sigma_{j_{1} j_{2}}} \right)^{2}
+
\left( \frac{m_{j_{3} j_{4}} - m_{Y}}{\sigma_{j_{3} j_{4}}} \right)^{2},
\end{equation}
where $m_{j_{k} j_{l}}$ denotes the invariant mass of the jet pair $(j_{k}, j_{l})$. 
The mass resolutions $\sigma_{j_{k} j_{l}}$ encode the expected off-shellness and are taken to be $30~\mathrm{GeV}$ for the $Z$ boson and $40~\mathrm{GeV}$ for the Higgs boson. A loop is performed over all possible jet permutations  and the configuration yielding the lowest $\chi^{2}_{XY}$ value is selected as the best reconstruction for that hypothesis. This minimization effectively identifies the jet pairs whose invariant masses lie closest to the true $Z$ or $h$ boson masses. From the selected combinations, we also determine the reconstructed transverse momenta $p_{T}$ and invariant masses of the $h$ and $Z$ candidates. \smallskip

The resulting $\chi^{2}_{XY}$ variables provide strong discriminating power. For  signal events, $\chi^{2}_{hh}$ values are typically lower compared to background processes. Similarly, for the $t\bar{t}ZZ$ ($t\bar{t}Zh$) backgrounds, the corresponding $\chi^{2}_{ZZ}$ ($\chi^{2}_{hZ}$) values are generally smaller than those of the signal or other backgrounds, aiding in the separation of the SM $\ttHH$ signal from competing processes. \smallskip

After studying these kinematic distributions, along with several others, we optimize a set of cuts on the relevant variables to suppress background events while retaining a significant portion of the signal, thereby maximizing the signal significance for the benchmark couplings. We find that events containing six jets, five of which are $b$-jets (i.e., the 6J5B category), yield the best significances, consistent with the SM case shown in Table~\ref{tab:sm-bsm-1l}. We then impose a series of hard cuts presented in Table~\ref{tab:cuts1l-HTbin}. After imposing the optimized cuts, we further exploit the  $H_T$ distribution by dividing it into eight  bins to enhance the significance even further.  Our choice for the $H_T$ binning is shown in Table~\ref{tab:cuts1l-HTbin}. \smallskip

\begin{table}[tb!]
 \centering
 \begin{tabular}{lr}\hline
Variable & Cuts \\ \hline
Leading jet $p_T$ & $p_T(j_1) \geq 100$ GeV \\
Second leading jet $p_T$ & $p_T(j_2)\geq 50$ GeV \\
Scalar $p_T$ sum of jets & $H_T\geq 400$ GeV\\
Scalar $p_T$ sum of b-tagged jets & $H_{Tb}\geq 300$ GeV\\
Invariant mass of jets &  $m_{\text{jets}}\geq 400$ GeV\\
Minimized-$\chi^2$ for $t\bar{t}hh$ topology & ${\rm Min}:\chi^2_{hh} <2$ \\
\begin{tabular}{c}
$p_T$ of leading reconstructed\\ Higgs  for $t\bar{t}hh$ topology
\end{tabular} & $p_T(h_1)^{hh}>250$ GeV\\
Sphericity, Aplanarity, $C$, $D$ & $\geq0.1$, $\geq0.02$, $\geq0.5$, $\geq0.1$ \\ \hline
$H_T$-bin edges & 400 - 600 - 800 - 900 - 1000 - 1200 - 1600 - 2000 - $\infty$ \\ \hline
\end{tabular}
 \caption{\label{tab:cuts1l-HTbin} Optimized cuts on the variables and binning of $H_T$ variable with events containing at least six jets ($N_j\geq 6$) of which five are $b$-jets ($N_b\geq 5$), i.e., the 6J5B category. } 
\end{table}

\begin{table}[h!]
\centering
\begin{tabular}{ccl}
\hline
Bin & $H_T$ Range [GeV] & Events\\
\hline
1 & $400$--$600$       & 257 \\
2 & $600$--$800$       & 2854 \\
3 & $800$--$900$       & 1664 \\
4 & $900$--$1000$      & 1130 \\
5 & $1000$--$1200$     & 2007 \\
6 & $1200$--$1600$     & 1423 \\
7 & $1600$--$2000$     & 481  \\
8 & $2000$--$\infty$   & 49   \\
\hline
\multicolumn{2}{l}{Total} & 9866 \\
\hline
\end{tabular}
\caption{\label{tab:bkg-HT-bin} Total background, including SM $\ttHH$, event yields in $H_T$ bins in the SL channel for 3 \abinv luminosity at LO in QCD.}
\end{table}

Table~\ref{tab:bkg-HT-bin} presents the  $H_T$ event distribution for the total background after applying the cuts  in Table~\ref{tab:cuts1l-HTbin}. Assuming that the number of observed events is the predicted one for the background, we performed one- and two-parameter fits for HEFT using the $\chi^2$ function:
\begin{equation}\label{eq:binchisq}
\chi^2(C_{\rm HEFT}) = \sum_{i\in {\rm Bin}} \frac{\left(N_{{\rm SM} + C_{\rm HEFT}}^i - N_{\rm SM}^i\right)^2}{N_B^i} \;.
\end{equation}
The $\chi^2$ distributions as a function of the HEFT Wilson coefficients  are shown in Fig.~\ref{fig:chi-1p} for the SL category with the blue dashed lines standing for the cut-based analysis. In this figure we  assumed an integrated luminosity of 3 \abinv and incorporated  NLO $k$-factors. The gray dashed horizontal line stands for the $\chi^2$ value corresponding to 95\% confidence level (CL) for a single-parameter fit. The corresponding 95\% CL allowed ranges for the HEFT couplings are listed in the second column of Table~\ref{tab:limit-one-param}. Furthermore, the 95\% CL contours  for simultaneous two-parameter variations are shown in  Fig.~\ref{fig:limit-2d-pmbdt}  with dashed blue lines. \smallskip

We do not perform this cut-based and binned analysis for the DL channel, as our goal here is only to obtain an indicative estimate of the allowed ranges for the HEFT couplings. The most stringent constraints are ultimately obtained from the multivariate analysis using the BDT method discussed in the next subsection. \smallskip


\subsection{Constraints from Parametric BDT analysis}
\label{subsec:pm-bdt}

In this subsection, we extract the constraints on the HEFT couplings using a parametric boosted decision tree  framework. The goal of this analysis is to exploit the discriminating power of a multi-class machine-learning classifier capable of interpolating smoothly across the multidimensional HEFT parameter space, and subsequently use its output to construct $\chi^2$ functions for the HEFT couplings. Below, we detail the classification strategy, training configuration, performance, and the procedure for extracting the resulting constraints. \smallskip

We perform a multi-class boosted decision tree classification using the \texttt{XGBoost} package~\cite{Chen_2016}. The classifier is trained on nine physics classes, consisting of the HEFT $\ttHH$ signal, represented through multiple benchmark samples, and eight background categories constructed from twelve background processes. The classes include
\begin{align*}
 {\rm HEFT}-\ttHH,\ \ttH,\ \ttZ,\ \ttb+4b,\ \ttb+2b2c,\ \ttb+2b,\ 
 \ttb\,+jets,\ \ttb SS,\ \ttb 4c4t,
\end{align*}
where $\ttb SS$ denotes the grouped backgrounds $\{\ttZH,\ \ttb VV,\ \ttb V+2b,\ \ttb h+2b\}$ and $\ttb 4c4t$ includes the four-top production ($\ttb\ttb$) and $\ttb+4c$. We also include the $c_{tg}$-induced  modifications to the $\ttH$ production, as in previous sections, and classify it as a HEFT contribution together with $\ttHH$, which increases the sensitivity to this coupling. To ensure robust statistics for the classifier training and testing, we restrict the input events to those satisfying the basic 4J3B selection, which provides a good balance between signal efficiency and available MC samples for all background classes. \smallskip

\subsubsection{Input variables}
\label{sec:inputs}
\begin{table}[!htb]
\centering
\renewcommand{\arraystretch}{1.5}
\footnotesize
\begin{tabular}{|l|l|}\hline
 Group & Variables \\ \hline
Object multiplicity & $N_j$,  $N_b$,   \\\hline
Transverse momenta  & $p_T(\ell)$,     $p_T(j_1)$,    $p_T(j_2)$,    $p_T(j_3)$,    $p_T(j_4)$,    \\\hline
Pseudo-rapidity & $\eta(\ell)$, $\eta(j_1)$,  $\eta(j_2)$,  $\eta(j_3)$,  $\eta(j_4)$,   \\\hline
Missing energy and Hadronic $p_T$, & $\mbox{\it MET}$,    $H_T$,    $H_{Tb}$,    $H_T+p_T(\ell)$,    \\\hline
Invariant masses & $m_{\rm jets}$,    $m_{{\rm jets}+\ell}$,    $m_T({\rm jets}+\ell)$,    $m_T(\mbox{\it MET}+{\rm jets}+\ell)$,    $m_{b{\rm s}}$,    $m_{b{\rm s}+\ell}$,     \\\hline
Conic separation & $\Delta  R(\ell,  j_1)$,  $\Delta  R(\ell,  j_2)$,  $\Delta  R(\ell,  j_3)$,  $\Delta  R(\ell,  j_4)$,   \\\hline
Azimuthal separation & \begin{tabular}{l} $\Delta\phi(\ell,  j_1)$,  $\Delta\phi(\ell,  j_2)$,  $\Delta\phi(\ell,  j_3)$,  $\Delta\phi(\ell,  j_4)$,   $\Delta\phi(\ell^+,\ell^-)$ - DL,\\
$\Delta\phi(\mbox{\it MET},  \ell)$,   
$\Delta\phi(\mbox{\it MET},  j_1)$,  $\Delta\phi(\mbox{\it MET},  j_2)$,\\  $\Delta\phi(\mbox{\it MET},  j_3)$,  $\Delta\phi(\mbox{\it MET},  j_4)$, \end{tabular}  \\\hline
Min, Max, average mass & ${\rm Min}:m_{bb}$,    ${\rm Max}:m_{bb}$,    ${\rm Avg}:m_{bb}$,    
${\rm Min}:m_{jj}$,    ${\rm Max}:m_{jj}$,    ${\rm Avg}:m_{jj}$,    \\\hline
Min, Max, average rapidity gap & \begin{tabular}{l}${\rm Min}:\Delta\eta_{bb}$,   ${\rm Max}:\Delta\eta_{bb}$,  ${\rm Avg}:\Delta\eta_{bb}$,  \\
${\rm Min}:\Delta\eta_{jj}$,   ${\rm Max}:\Delta\eta_{jj}$,   ${\rm Avg}:\Delta\eta_{jj}$,  \end{tabular}\\\hline
Min, Max, average conic separation & \begin{tabular}{l}${\rm Min}:\Delta R(bb)$,   ${\rm Max}:\Delta R(bb)$,   ${\rm Avg}:\Delta R(bb)$, \\
${\rm Min}:\Delta R(jj)$,   ${\rm Max}:\Delta R(jj)$,   ${\rm Avg}:\Delta R(jj)$,  \end{tabular} \\\hline
Minimized $\chi^2$ for $hh$, $hZ$, $ZZ$ topology &${\rm Min}:\chi^2_{hh}$,  ${\rm Min}:\chi^2_{hZ}$,  ${\rm Min}:\chi^2_{ZZ}$,  \\\hline
Reconstructed variables & \begin{tabular}{l}$m_{h_1}^{hh}$,    $p_T(h_1)^{hh}$,    $m_{h_2}^{hh}$,    $p_T(h_2)^{hh}$,    \\
$m_h^{hZ}$,    $p_T(h)^{hZ}$,    $m_Z^{hZ}$,    $p_T(Z)^{hZ}$,    \\
$m_{Z_1}^{ZZ}$,    $p_T(Z_1)^{ZZ}$,    $m_{Z_2}^{ZZ}$,    $p_T(Z_2)^{ZZ}$,    \\
$\Delta\phi(\ell,  h_1)$,  $\Delta\phi(\ell,  h_2)$,  $\Delta\phi(h_1,  h_2)$, \end{tabular}\\\hline
Event shape variables & ${\rm Sphericity}$,  ${\rm Aplanarity}$,  $C$,  $ D$ . \\ \hline
\end{tabular}
\caption{\label{tab:inputvariables} List of variables used in BDT training.}
\end{table}

For all events that satisfy the baseline selection (4J3B), a comprehensive set of kinematic and topological observables is constructed and used as input to the BDT classifier. These variables, summarized in Table~\ref{tab:inputvariables}, are chosen to capture characteristic differences between the HEFT signal and background processes and to enhance the separation power of the multivariate analysis.

\begin{itemize}
\item
The object multiplicity variables include the total jet multiplicity $N_j$ and the number of $b$-tagged jets $N_b$. These quantities provide coarse information about the underlying partonic activity and are particularly useful for distinguishing $\ttHH$ from backgrounds with fewer heavy-flavour jets.
\item
The transverse momentum observables, $p_T(\ell)$ and $p_T(j_{1,2,3,4})$, encode the hardness of the event. Since the signal typically contains multiple energetic jets from top and Higgs decays, these variables contribute significantly to the discrimination.
\item
The pseudorapidity variables, $\eta(\ell)$ and $\eta(j_{1,2,3,4})$, characterize the event topology in the detector and help differentiate central, signal-like events from more forward, background-like configurations.
\item
Global activity variables such as the missing transverse momentum ($\mbox{\it MET\,}$) and the hadronic scalar sums ($H_T$, $H_{Tb}$, and $H_T + p_T(\ell)$) quantify the inclusive momentum flow in the event. These are effective in separating backgrounds with softer hadronic components or less missing energy.
\item
Several invariant-mass and transverse-mass observables, $m_{\rm jets}$, $m_{{\rm jets}+\ell}$, $m_T({\rm jets}+\ell)$, $m_T(\mbox{\it MET}+{\rm jets}+\ell)$, $m_{b{\rm s}}$, and $m_{b{\rm s}+\ell}$, probe combinations of jets and leptons that are expected to reconstruct intermediate resonances in the signal.
\item
The angular-separation variables include conic separations $\Delta R(\ell, j_i)$ and azimuthal differences $\Delta\phi(\ell, j_i)$, $\Delta\phi(\ell^+,\ell^-)$ in DL channel  as well as their analogues involving the missing energy. These quantify the relative orientation among visible objects and between visible objects and $\mbox{\it MET}$, providing sensitivity to the distinct decay patterns of signal and background.
\item
To capture event-level correlations among jets and $b$-jets, we include the minimum, maximum, and average values of $m_{bb}$, $m_{jj}$, $\Delta\eta_{bb}$, $\Delta\eta_{jj}$, $\Delta R(bb)$, and $\Delta R(jj)$. These aggregate shape descriptors summarize the internal structure of the hadronic system.
\item
A particularly powerful set of inputs consists of the minimized $\chi^2_{XY}$ variables for the three possible topologies: $hh$, $hZ$, and $ZZ$, discussed above. For each hypothesis, all possible jet  assignments are tested, and a $\chi^2$ value is computed to quantify the compatibility with the target topology. The minimum over all permutations, ${\rm Min}:\chi^2_{hh}$, ${\rm Min}:\chi^2_{hZ}$, and ${\rm Min}:\chi^2_{ZZ}$, serves as a strong discriminator: signal events preferentially yield smaller $\chi^2_{hh}$, while background processes such as $\ttZZ$ or $\ttZH$ tend to minimize the corresponding $\chi^2$ for their respective topologies.
\item
The analysis also uses reconstructed Higgs and $Z$ candidates identified by the combinations yielding the minimum $\chi^2$ in each topology. Their reconstructed masses and transverse momenta, $m_{h_1}^{hh}$, $p_T(h_1)^{hh}$, $m_{h_2}^{hh}$, $p_T(h_2)^{hh}$, $m_h^{hZ}$, $m_Z^{hZ}$, $m_{Z_1}^{ZZ}$, $m_{Z_2}^{ZZ}$, and the corresponding $p_T$ values, provide additional resonant structure information. Angular differences involving these reconstructed objects, such as $\Delta\phi(\ell, h_{1,2})$ and $\Delta\phi(h_1, h_2)$, further refine the discrimination in events where the Higgs bosons are moderately boosted.
\item
Finally, event-shape variables including the sphericity, aplanarity, and the $C$ and $D$ parameters summarize the overall geometry of the final state and help distinguish the more isotropic signal events from the typically more planar QCD-dominated backgrounds~\cite{Dasgupta:2003iq,Banfi:2010xy,Weber:2009bhh}.
\end{itemize}

\begin{figure}[tb!]	
\centering	
\begin{minipage}{0.43\textwidth}
\centering
\includegraphics[width=1\textwidth]{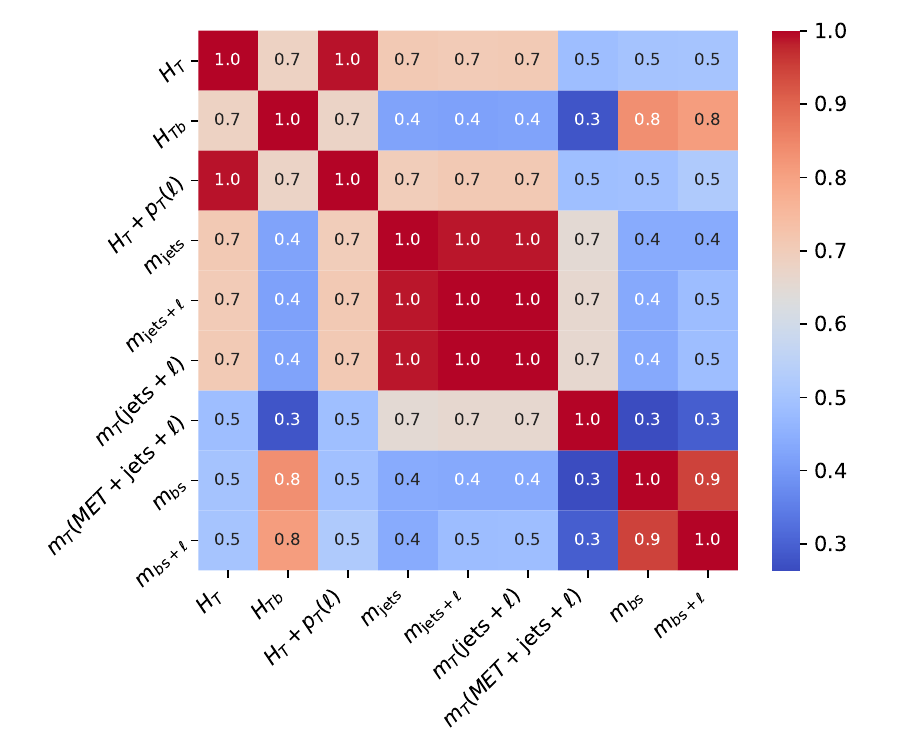}
\includegraphics[width=1\textwidth]{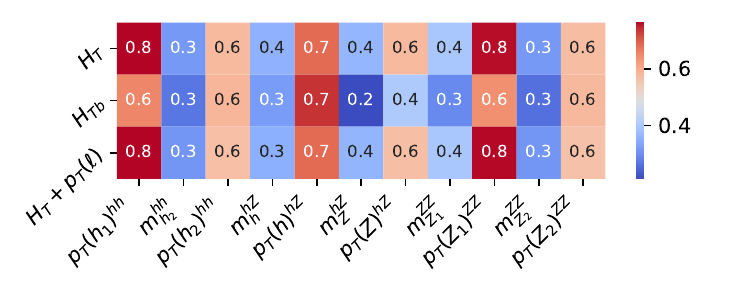}
\end{minipage}
\hfill	
\begin{minipage}{0.56\textwidth}
\includegraphics[width=1\textwidth]{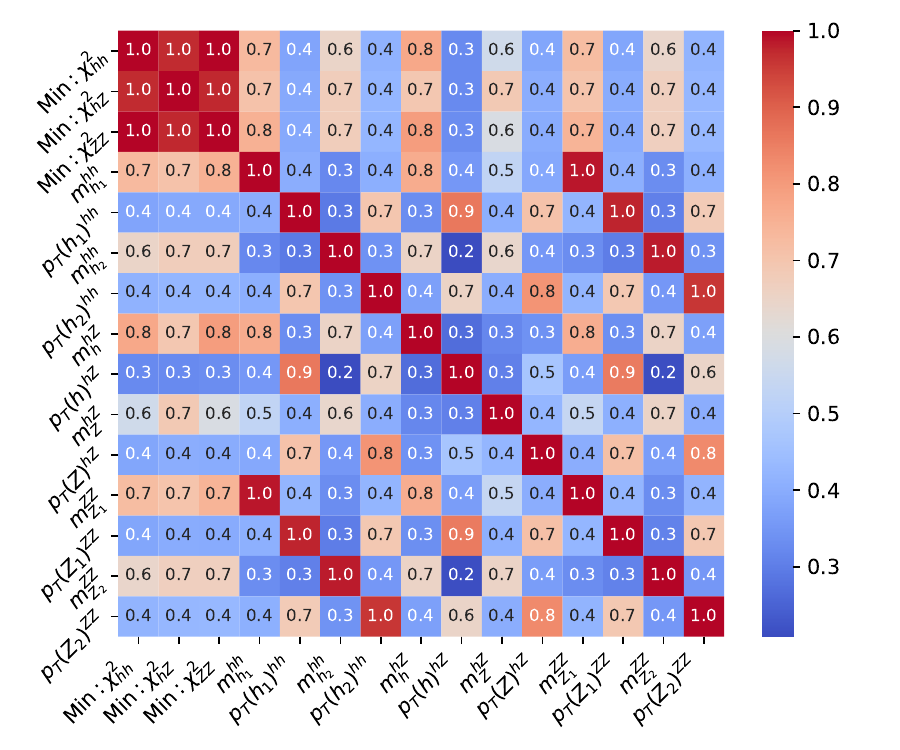}	
\end{minipage}
\caption{\label{fig:var-corr} Correlation among the global event variables, reconstructed di-boson topological variables, and global vs. the topological variables are presented for the $\ttHH$ process with HEFT coupling $c_2=-2.5$.}
\end{figure}

All these variables are, however, not independent of one another, as expected. For example, the global activity variables such as $\mbox{\it MET}$, $H_T$, and the invariant masses exhibit mutual correlations. The reconstructed di-boson topological variables, such as the transverse momenta of the reconstructed Higgs and $Z$ candidates, also show strong internal correlations. Reasonable correlations are further observed between the global variables and the reconstructed di-boson observables. The event-shape variables are highly correlated among themselves, while the $\Delta\eta$ and $\Delta R$ variables display the expected correlations due to their geometric relationship. Representative examples of these correlations, including those among global event variables, reconstructed di-boson topological variables, and between global and topological observables, are shown in Fig.~\ref{fig:var-corr} for the $\ttHH$ process with the HEFT coupling choice $c_2 = -2.5$. \smallskip

Together, these variables constitute a high-dimensional feature set that captures both local (object-level) and global (event-level) information. This enables the BDT to efficiently discriminate between signal and background over the full kinematic range relevant for extracting the HEFT-induced $\ttHH$ signal while implicitly accounting for correlations among the input variables. \smallskip

%
\subsubsection{Parametric BDT methodology}
To extract constraints on the HEFT couplings, we employ a parametric BDT, in which the values of the HEFT couplings are provided as additional input features, along with the usual kinematic variables. Multiple $\ttHH$ samples generated at different HEFT benchmark points are merged into the single signal class, while the numerical values of the couplings are appended as extra nodes in the feature space. This enables the BDT to interpolate smoothly across the parameter space and allows the extraction of continuous constraints from discrete benchmark samples. A detailed discussion of parametric machine-learning approaches in high-energy physics analysis can be found in Refs.~\cite{Baldi:2016fzo,Chatterjee:2021nms}.\smallskip

To prevent the classifier from trivially distinguishing signal and background via the coupling node, since background events have no physical coupling assignment, we randomly assign to each background event a coupling value drawn from the same set of benchmark points. This ensures that the coupling information alone cannot serve as a discriminator. 
We perform the analysis one-parameter at a time (total   five classifiers) and also two parameter at a time (a total seven classifiers) with the bench mark couplings discussed before. 
To obtain stable and unbiased predictions, we optimize training hyperparameters and adopt conservative regularization to prevent over-fitting. \smallskip

\begin{figure}[tb!]
 \centering 
\includegraphics[width=0.49\textwidth]{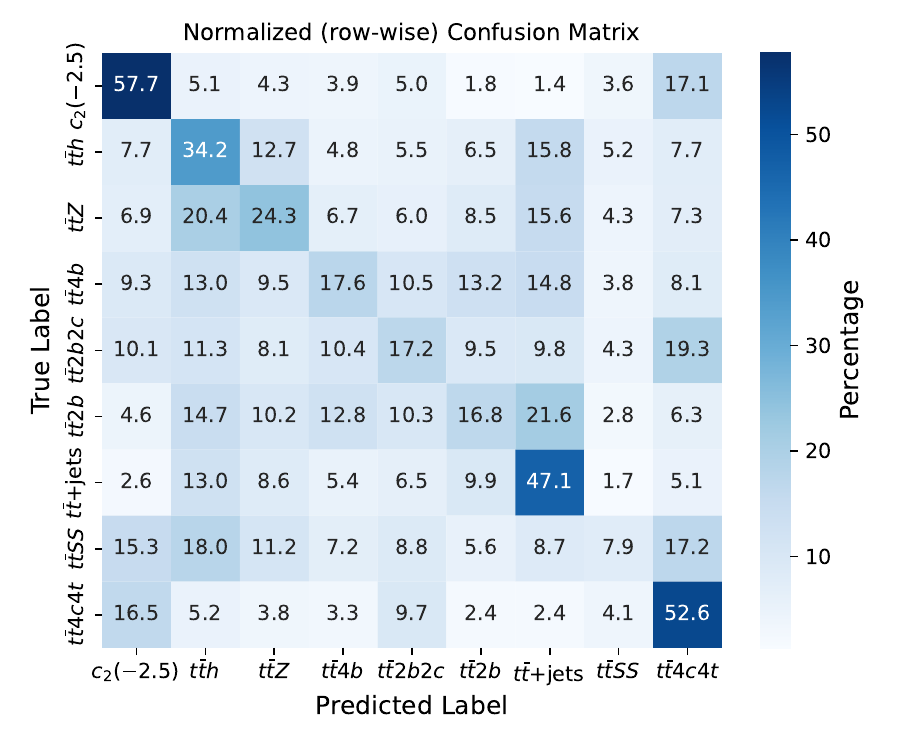} 
 \includegraphics[width=0.49\textwidth]{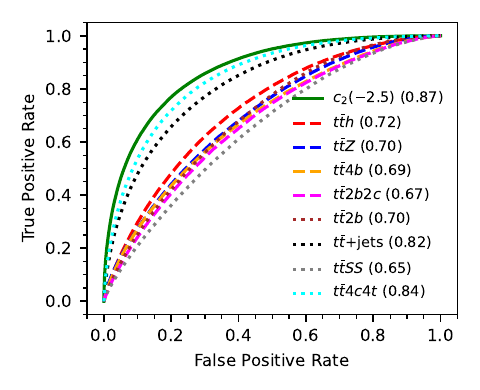}
 \includegraphics[width=0.49\textwidth]{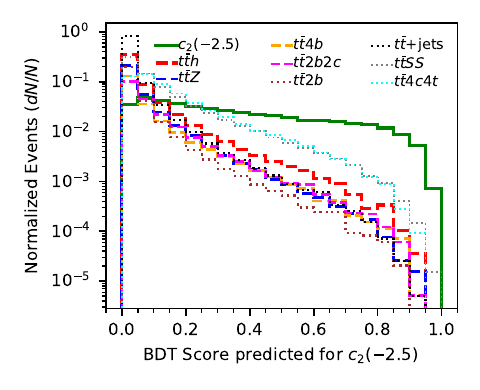} 
 \includegraphics[width=0.49\textwidth]{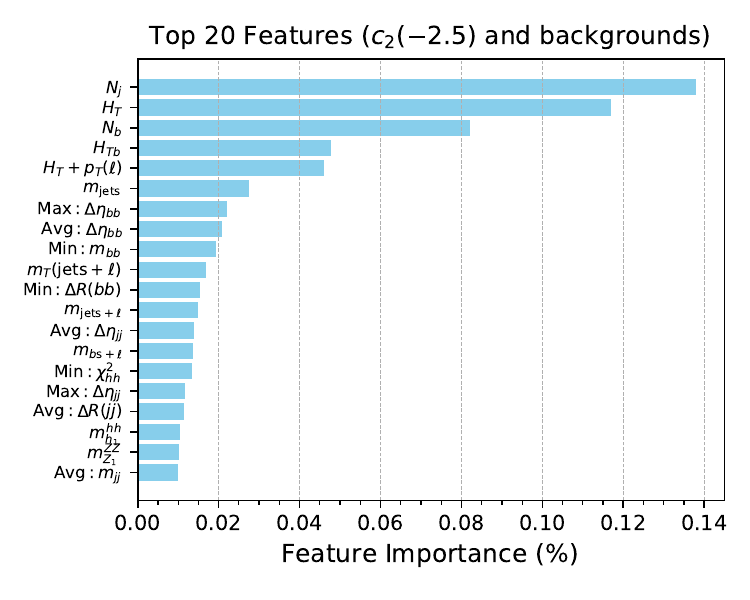}
 \caption{\label{fig:bdt-bsm-bkg} Confusion matrix (left-top), ROC curves (right-top), BDT score for HEFT $\ttHH$ (left-bottom), and top twenty features (right-bottom) from multi-class BDT training of $c_2=-2.5$ HEFT couplings as an example are shown. The numbers inside the parenthesis with the process names on the ROC curves corresponds to area under the corresponding ROC curves.}
\end{figure}

An illustration of the classification performance, for example, at $c_2=-2.5$, is shown in Fig.~\ref{fig:bdt-bsm-bkg} through the confusion matrix, ROC curves,  the HEFT-$\ttHH$ BDT score distribution, and the ranking of the top 20 discriminating features for the SL channel. The PM-BDT achieves a clear separation between the signal and the background categories. From the confusion matrix, the background least similar to the HEFT $\ttHH$ signal is the $\mathrm{t\bar{t}}$+jets, which is misclassified as $\ttHH$ at the 2.6\% level, while the background with the largest confusion into $\ttHH$ is $tt4c4t$ with 16.5\%, followed by $t\bar{t}SS$ with 15.3\%.\smallskip

Conversely, the HEFT $\ttHH$ itself, followed by the $tt4c4t$ and $\mathrm{t\bar{t}}$+jets classes, shows the strongest self-identification classification performance. Classes such as $\ttH$, $\ttZ$, and $\mathrm{t\bar{t}+2b}$ exhibit moderate but well-behaved separability, consistent with their kinematic proximity to the signal.  The ROC curves further corroborate the trends seen in the confusion matrix. The HEFT-induced $\ttHH$ class achieves the highest discrimination with an area under the curve (AUC) of 0.87 (shown in parentheses in the figure). The next classes are  $tt4c4t$ and $\mathrm{t\bar{t}}$+jets  with AUCs of 0.84 and 0.83, respectively, while $\ttH$ yields an AUC of 0.72. \smallskip

The HEFT $\ttHH$ BDT score distribution, shown in the lower-left panel, is consistent with the trends observed in the confusion matrix. The signal distribution remains relatively broad and flat in logarithmic scale, indicating a stable separation from background across the full score range. Among the backgrounds, $\ttb 4c4t$ and $\ttb SS$ exhibit shapes most similar to the signal and populate the intermediate-score region, while $\mathrm{t\bar{t}}+$jets, being the least signal-like, peaks sharply near zero and falls rapidly even on a logarithmic scale.\smallskip

The ranking of the top 20 discriminating features indicates that global event activity and heavy-flavor multiplicity dominate the classification. The number of jets $N_j$ is the single most powerful variable ($\sim 14\%$), followed by the total scalar transverse energy $H_T$ ($\sim 12\%$) and the number of $b$-tagged jets $N_b$ ($\sim 8\%$). Together, these observables encode the high jet multiplicity, large hadronic activity, and enhanced heavy-flavor content characteristic of the $\ttHH$ final state. \smallskip

Subleading but still significant contributions arise from $H_T$-related observables, including $H_{Tb}$ and $H_T + p_T(\ell)$, which are highly correlated with $H_T$ and probe the overall hardness of the event. Invariant-mass-based variables, such as the total jet mass and jet--lepton mass combinations, further enhance discrimination by reflecting the presence of multiple heavy resonances in the final state. Additional separation power is provided by angular correlations among $b$-jet and light-jet pairs, which are sensitive to Higgs decay topologies and multi-jet radiation patterns. Variables related to partial Higgs reconstruction, including $\chi^2_{hh}$ and reconstructed Higgs masses, retain discriminating power despite combinatorial ambiguities, while observables such as $m_{Z_1}^{ZZ}$ and averaged dijet masses contribute at the few-percent level ($\sim 1\%$), representing the least discriminating features among the top-ranked variables. Notice that the leading discriminators are in line with our cut-based analysis, as expected. \smallskip

After training, we obtain the classifier probability scores for the HEFT $\ttHH$ class for all benchmark points as well as for all background processes. The signal and background events are binned into ten regions based on the BDT output variable. In each bin, we evaluate the expected signal and background yields and construct a $\chi^2$ in analogy to Eq.~(\ref{eq:binchisq}). 
The one-parameter $\chi^2$ functions as a function of the HEFT couplings are shown in Fig.~\ref{fig:chi-1p} for both the SL and DL channels, denoted by the solid dark-green and dashed–dotted red curves, respectively, with the corresponding cut-based $\chi^2$ results obtained earlier.  The combined $\chi^2$ obtained from the SL and DL channels is also shown in the same figure using dotted magenta lines. The one-parameter 95\% C.L. intervals are shown in Table~\ref{tab:limit-one-param} for the SL, DL channels and their combination. \smallskip

The corresponding allowed ranges for the HEFT couplings under simultaneous two-parameter variations are shown in Fig.~\ref{fig:limit-2d-pmbdt}. The 95\% C.L. contours (solid dark-green)
are obtained from the fitted two-parameter $\chi^2$ functions. These results are presented only for the SL channel in order to illustrate the parameter correlations and their interplay. For comparison, the contours from the cut-based analysis are also shown with dashed blue lines.\smallskip

\section{Results and Interpretations}
\label{sec:results}

This section summarizes the main results obtained from the different analysis strategies considered in this study, namely the cut-based approach and the PM:BDT  method, as well as  it provides a coherent interpretation of the corresponding sensitivities. We compare the performance of the cut-based $H_T$-binned analysis in the single-lepton channel with the PM:BDT results in the SL and dilepton channels, as well as their statistical combination. The emphasis is on understanding the relative improvement achieved through multivariate techniques and channel combination  and on positioning our most stringent limits in the context of existing experimental constraints and future prospects.\smallskip

\begin{figure}[htb!]
	\centering	
		\includegraphics[width=1\textwidth]{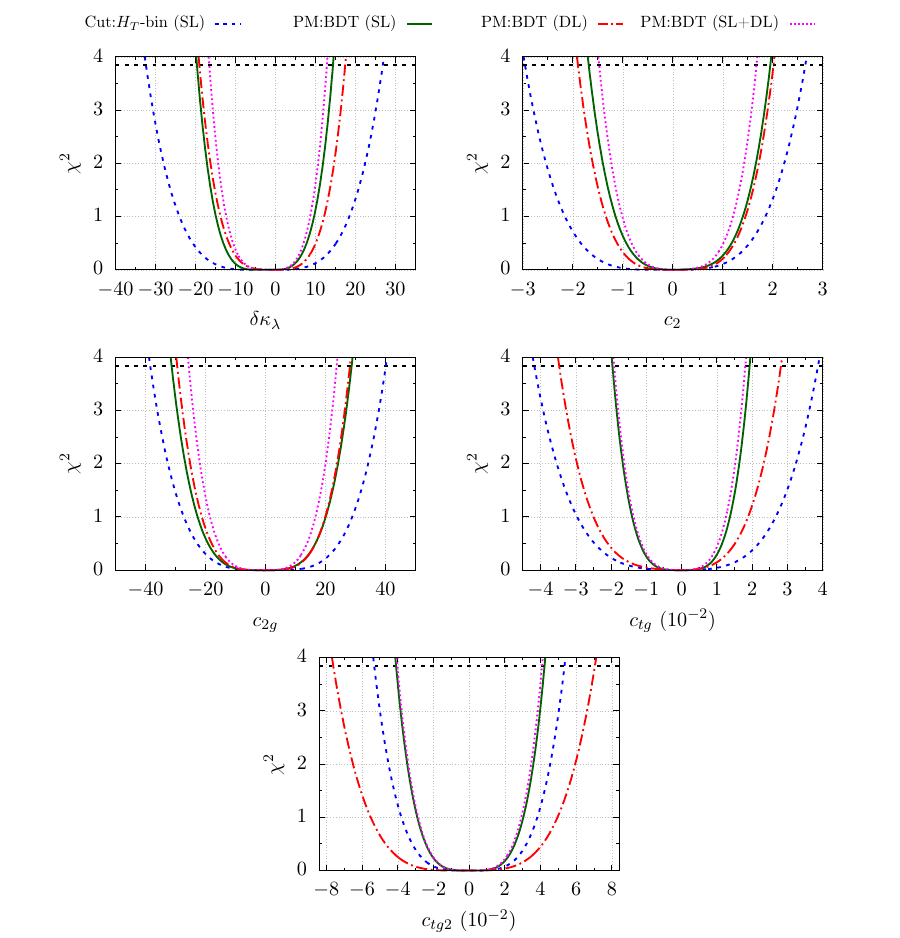}
	\caption{\label{fig:chi-1p} 
		The $\chi^2$ as a function of the HEFT Wilson coefficients for the cut-based analysis (Cut:$H_T$-bin) in the SL channel, the PM:BDT analysis in the SL and DL channels, and the combined SL+DL PM:BDT analysis are shown for an integrated luminosity of 3 ab$^{-1}$. The horizontal gray dashed line 
    indicates the 95\% C.L. interval on the HEFT couplings. }
\end{figure}

From the 95\% CL limits presented in Table~\ref{tab:limit-one-param} and the one-parameter $\chi^2$ distributions shown in Fig.~\ref{fig:chi-1p}, we observe that the PM:BDT approach in the single-lepton (SL) channel significantly improves the sensitivity compared to the cut-based $H_T$-binned SL analysis for all HEFT couplings, leading to visibly tighter confidence intervals. 

\begin{table}[tb!]	
	\centering
	\renewcommand{\arraystretch}{1.2}
	\begin{tabular}{lcccc}\hline
Couplings & Cut:$H_T$-bin (SL)   & PM:BDT (SL) & PM:BDT (DL) & PM:BDT (SL+DL) \\ \hline
$\delta\kappa_\lambda   $&[-32.4,~+26.8]&[-19.6,~+14.4]&[-19.0 ,~+17.4 ]&  [-16.5 ,~+12.9 ]   \\ \hline
$c_2$                   &[-2.95,~+2.65   ]&[-1.68,~+1.95 ]&[-1.9  ,~+2.02 ]&  [-1.47 ,~+1.68 ]   \\ \hline
$c_{2g}$                &[-38.4,~+40.1 ]&[-31.3,~+28.7]&[-29.4 ,~+28.2 ]&  [-25.6 ,~+23.7 ]   \\ \hline
$c_{tg}$                &[-0.042,~+0.039   ]&[-0.02,~+0.019 ]&[-0.035,~+0.028]&  [-0.019,~+0.018]   \\ \hline
$c_{tg2}$               &[-0.053,~+0.053] &[-0.041,~+0.042]&[-0.078,~+0.072]& [-0.040,~+0.041]\\ \hline
\end{tabular}
\caption{\label{tab:limit-one-param} One parameter limits are shown on HEFT couplings  for cut-based (Cut:$H_T$-bin) in SL channel and PM:BDT approach in both SL and DL channel at 3000 fb$^{-1}$ luminosity without systematic uncertainty. The combined limits from SL and DL channel are also shown for PM:BDT case (best) in the last column.  The corresponding $\chi^2$ plots are shown in Fig.~\ref{fig:chi-1p}.}
\end{table}

\begin{figure}[htb!]
	\centering	
	\includegraphics[width=1\textwidth]{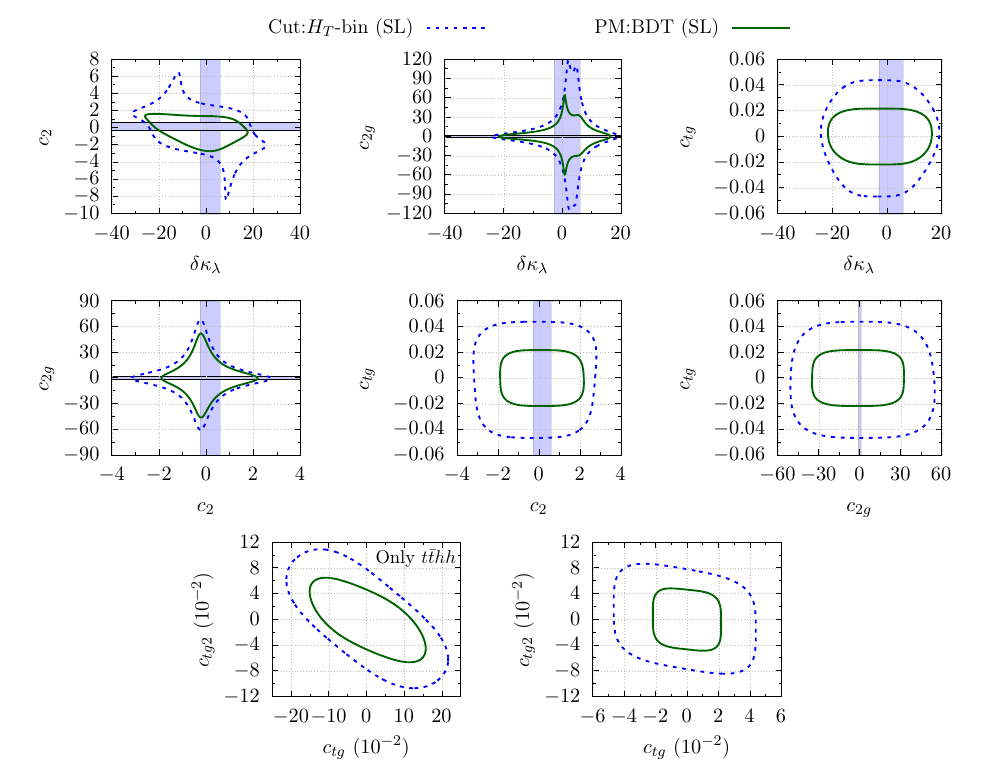}
	\caption{\label{fig:limit-2d-pmbdt} The 95\% C.L. contours ($\chi^2 = 5.99$) showing the allowed regions of the HEFT couplings obtained using a cut-based analysis (Cut:$H_T$-bin) and a parametric BDT analysis  in the single-lepton (SL) channel are shown for an integrated luminosity of $3~\mathrm{ab}^{-1}$. The light-blue bands for $\delta\kappa_\lambda$  represents the available experimental 95\% C.L. allowed ranges on the couplings~\cite{ATLAS:2025hhd}; see the text for details.}
\end{figure}

The dilepton  PM:BDT limits are in general comparable to, though slightly weaker than, the SL PM:BDT results, reflecting the reduced statistics in the DL channel despite its cleaner final state. Nevertheless, the DL channel provides complementary sensitivity, especially for the coupling $c_{2g}$, for which the limits are slightly stronger in the DL channel, driven by the comparatively higher sensitivity of $\Delta\phi(\ell^+,\ell^-)$ to $c_{2g}$, see Fig.~\ref{fig:dist-bsm}. Furthermore the best constraints are obtained by combining the SL and DL channels within the PM:BDT framework. The PM:BDT (SL+DL) limits are consistently the most stringent, yielding an additional improvement of about 10–20\% over the individual channels. This demonstrates the clear advantage of multivariate techniques and channel combination in maximizing sensitivity to HEFT couplings at high luminosity. \smallskip

Let us now discuss the strongest results PM-BDT(SL+DL) limits quoted in Table~\ref{tab:limit-one-param} in view of the existing experimental ATLAS and CMS results which correspond either to the total Run 2 integrated luminosity or at best include the first 3 years of Run 3, thus corresponding to about 5–10\% of the total expected luminosity for HL-LHC. We note that the current ATLAS and CMS results are going to be soon surpassed by higher luminosity, higher recorded data and analysis tools quality of the Run 3 at LHC.
ATLAS experiment~\cite{ATLAS:2025hhd} constrains the Higgs self-coupling modifier to $\delta\kappa_\lambda \in [-2.6,~5.6]$ at 95\% C.L., using the $b\bar b\gamma\gamma$ channel with 308~fb$^{-1}$, which is  already stronger than our $t\bar t hh$ constraint of $\delta\kappa_\lambda \in [-16.5 ,~12.9]$ obtained at 3~ab$^{-1}$.

The ATLAS~\cite{ATLAS:2024ish} and CMS~\cite{CMS:2025ngq} experiments constrain the Higgs contact interaction coupling $c_2$ from the ggF $hh$ process to $c_2 \in [-0.19,~0.7]$ with 140~fb$^{-1}$ and $c_2 \in [-0.28,~0.59]$ with 138~fb$^{-1}$, respectively. ATLAS~\cite{ATLAS:2024ish} also constrains the gluon–Higgs contact interaction coupling $c_{2g}$ from ggF($hh$) analyses, yielding $c_{2g} \in [-1.47,~1.14]$.
These results are obtained by both experiments by combining several detailed ggF($hh$) analyses, each covering a wide range of possible Higgs decay channels.
A recent ATLAS analysis~\cite{ATLAS:2026dvm}, using the $t\bar{t}hh$ process with 196~fb$^{-1}$ and covering a subset of decay channels of both the top quark and the Higgs boson, yields $c_2 \in [-3.9,~3.3]$. This process is also under study at CMS~\cite{CMS:2025jzc}.
Comparing these experimental results on $c_2$ with each other, and moreover with those obtained in this work ($c_2 \in [-1.47, ~1.68]$, $c_{2g} \in [-25.6, ~23.7]$) under very different conditions, should be done with great caution from both experimental and theoretical perspectives. Nevertheless, for an integrated luminosity of 196~fb$^{-1}$, our projected limits on $c_2$ are slightly stronger than the current ATLAS bounds.

This  difference of limits on $\delta\kappa_\lambda$, $c_2$ and $c_{2g}$ arises from the much smaller production cross section and reduced signal efficiency associated with high-multiplicity final states in the $t\bar t hh$ channel.

Nevertheless, it is important to emphasize that the $t\bar{t}hh$ channel provides a complementary probe of these interactions. In ggF production, $c_2$ and $c_{2g}$ enter together with additional effective couplings, and the resulting limits are typically derived under simplifying assumptions by fixing other couplings to their SM values. Relaxing these assumptions requires a multidimensional fit, which can weaken the extracted bounds. In contrast, $t\bar{t}hh$ probes these interactions in a different topology with reduced degeneracies, thereby offering an independent handle on the underlying HEFT structure.

In contrast, there are currently no experimental bounds on the HEFT coefficients $c_{tg}$ and $c_{tg2}$, which parameterize the $g(g)t\bar{t}h$ and $g(g)t\bar{t}hh$ interactions, respectively.
\smallskip

 Although the one-dimensional constraint on $\delta\kappa_\lambda$ 
 is weaker than current experimental limits, its inclusion in the two-dimensional analysis is well motivated, since variations in this coupling significantly reshape the allowed regions of the other HEFT parameters. The two-parameter allowed regions at 95\% CL are shown in Fig.~\ref{fig:limit-2d-pmbdt}.  As expected, the cut-based SL contours are considerably broader than those obtained with the PM:BDT approach in the SL channel, highlighting the improved sensitivity of the multivariate framework.\smallskip

Focusing on the PM:BDT SL contours, a clear pattern emerges when $\delta\kappa_\lambda$ is varied simultaneously with other couplings. Once $\delta\kappa_\lambda$ deviates from zero, the allowed regions for $c_2$ and $c_{2g}$ shrink noticeably, indicating that nonzero Higgs self-coupling modifications enhance sensitivity to operators affecting both the top–Higgs interaction and effective gluonic couplings. This behavior can be traced to the strong impact of $\delta\kappa_\lambda$ on the di-Higgs kinematics, which in turn constrains additional contributions that distort similar phase-space regions. The contours in the $c_2$–$c_{2g}$ plane exhibit a rather intricate correlation, whereby a deviation of either coupling from zero significantly restricts the allowed range of the other, reflecting their mutually constraining interplay. 
\smallskip

In contrast, the correlations of $c_{tg}$ with the other couplings, except for $c_{tg2}$, remain relatively weak. The allowed range of $c_{tg}$ exhibits only marginal variations when the remaining parameters are varied, indicating a limited interplay between this operator and the rest of the parameter space within the PM:BDT SL analysis. This is mainly due to the inclusion of $c_{tg}$ effects in $\ttH$ production in addition to $\ttHH$. When neglecting the $\ttH$ contribution to $c_{tg}$, we observe that $c_{tg}$ and $c_{tg2}$ exhibit a moderate negative correlation (left-bottom panel) owing to their similar vertex structure. However, this correlation is partially reduced once the $\ttH$ contribution to $c_{tg}$ is included, as the latter process provides an additional independent constraint.\smallskip

In summary, the results presented in this section demonstrate that multivariate parametric learning substantially enhances the sensitivity to HEFT couplings in the challenging $t\bar t hh$ final state at the HL-LHC. The PM:BDT approach consistently outperforms the cut-based analysis, while the combination of single-lepton and dilepton channels yields the strongest projected constraints. Although the one-dimensional limit on $\delta\kappa_\lambda$ is weaker than current experimental bounds, its inclusion plays a crucial role in constraining the multidimensional parameter space through nontrivial correlations with other HEFT operators. The two-parameter analyses further reveal characteristic interference patterns and degeneracy lifting among the couplings, highlighting the complementary information encoded in correlated observables. In addition, to validate the EFT treatment of the HEFT benchmark choices adopted in our analysis, as discussed in Sect.~\ref{subsec:HEFT-samples}, we verified that imposing an upper cut of $m_{t\bar{t}hh}<3$ TeV changes the projected sensitivities only marginally. These findings underline the importance of the interplay between $t\bar t hh$ and different Higgs production processes such as $h$, $hh$ and $\ttH$ to extract the HEFT couplings and fully exploit the SM and BSM physics potential of $\ttHH$ process at High luminosity.

\section*{Acknowledgements}

OJPE is partially supported by CNPq grant number 302120/2025-4 and FAPESP grants 2019/04837-9. RR is supported by FAPESP fellowship with grant 2023/04036-1 and 2025/06648-0. RDM is partially supported by FAPESP grants 2018/25225-9 and 2021/14335-0.
\appendix
\section{Two-parameter dependence of production cross section}\label{sec:sig-2param}
For completeness, we present here the full  expression of the $\ttHH$ production cross section containing all mixed contributions involving pairs of HEFT couplings. Terms linear in a single coupling, which were already introduced in Eq.~(\ref{eq:prod-cs-HEFT}), are therefore not repeated here:
\begin{align}\label{eq:sig-2param}
&\sigma_{\ttHH}(c_i,c_j) = 
 0.384 \times \delta\kappa_\lambda \times c_2 \nonumber\\[0.4em]
& + 0.0321 \times \delta\kappa_\lambda \times c_{2g}
- 3.19 \times 10^{-3} \times \delta\kappa_\lambda \times c_{2g}^2
+ 4.93 \times 10^{-4} \times \delta\kappa_\lambda^2 \times c_{2g}^2
\nonumber\\[0.4em]
& + 8.16 \times 10^{-3} \times \delta\kappa_\lambda \times c_{tg}
- 6.49 \times 10^{-4} \times \delta\kappa_\lambda^2 \times c_{tg}
+ 1.83\times10^{2} \times \delta\kappa_\lambda \times c_{tg}^2
+ 25.6 \times \delta\kappa_\lambda^2 \times c_{tg}^2
\nonumber\\[0.4em]
& + 0.0396 \times c_2 \times c_{2g}
+ 0.0135 \times c_2^2 \times c_{2g}
+ 9.97 \times 10^{-3} \times c_2 \times c_{2g}^2
+ 0.0163 \times c_2^2 \times c_{2g}^2
\nonumber\\[0.4em]
& - 6.57 \times 10^{-3} \times c_2 \times c_{tg}
+ 11.9 \times c_2 \times c_{tg}^2
\nonumber\\[0.4em]
& + 1.83 \times 10^{-3} \times c_{2g} \times c_{tg}
- 0.0249 \times c_{2g} \times c_{tg}^2\nonumber\\[0.4em]
& + 1848 \times c_{tg}\times c_{tg2}  - 0.973 \times c_{tg}^2\times c_{tg2}~.
\end{align}

\bibliographystyle{utphysM}
\bibliography{refsrafi,refs}
\end{document}